\documentclass[aps,prb,unsortedaddress,showpacs]{revtex4}

\usepackage{amsmath}
\usepackage{amssymb}
\usepackage{graphicx}
\usepackage{epsfig}

\begin{document}

\title{Quantum phase space function formulation of reactive flux theory}

\author{Debashis Barik}
\email{pcdb4@mahendra.iacs.res.in}
\affiliation{Indian Association for the Cultivation of Science, Jadavpur,
Kolkata 700 032, India}

\author{Suman Kumar Banik}
\email{banik@mpipks-dresden.mpg.de}
\affiliation{
Max-Planck-Institut f\"ur Physik komplexer Systeme, N\"othnitzer
Str. 38, 01187 Dresden, Germany}

\author{Deb Shankar Ray}
\email{pcdsr@mahendra.iacs.res.in}
\affiliation{Indian Association for the Cultivation of Science, Jadavpur,
Kolkata 700 032, India}

\date{\today}

\begin{abstract}
On the basis of a coherent state representation of quantum noise
operator and an ensemble averaging procedure a scheme for quantum
Brownian motion has been proposed recently [Banerjee {\it et al},
Phys. Rev. E {\bf65}, 021109 (2002); {\bf66}, 051105 (2002)]. We
extend this approach to formulate reactive flux theory in terms
of quantum phase space distribution functions and to derive a
time dependent quantum transmission coefficient - a quantum
analogue of classical Kramers'-Grote-Hynes coefficient in the
spirit of Kohen and Tannor's classical formulation. The theory is
valid for arbitrary noise correlation and temperature. The
specific forms of this coefficient in the Markovian as well as in
the non-Markovian limits have been worked out in detail for
intermediate to strong damping regime with an analysis of quantum
effects. While the classical transmission coefficient is
independent of temperature, its quantum counterpart has
significant temperature dependence particularly in the low
temperature regime.
\end{abstract}


\maketitle

\section{Introduction}
The classic work of Kramers \cite{kramer} on the diffusion model
of chemical reactions in 1940 forms the dynamical basis of modern
rate theory of activated processes. With the development of
ultrafast lasers and time resolved detection techniques since mid
seventies it has been possible to monitor the details of the
pathways of a reaction on a microscopic scale \cite{dian,caste}.
This has provided a major impetus and also generated a renewal of
interest in theoretical development in chemical dynamics,
particularly in the problem of reactions in condensed media. Over
the decades the field has grown in various new directions, e.g. ,
extension of Kramers results to non-Markovian regime
\cite{{grote-hynes},{hang-mojta},car}, generalizations to higher
dimensions \cite{lan,pollak}, inclusion of complex potentials
\cite{gram1,gram2}, generalization to open systems
\cite{ray1,ray2,ray3}, analysis of semiclassical and quantum
effects
\cite{ray4,weiss,grab,hangg,woly,mill,calde,grab1,topa,bern,
db1,skb,db2,db3,bic1,sch},
thermal ratchet \cite{mag} and molecular motors \cite{juli} etc.
These developments have been the subject of several reviews and
monographs.We refer to \cite{weiss,grab,hangg,topa}.

It is now well known that the rate constant of a chemical
reaction may be realized as a typical transport coefficient in
phase space. This transport coefficient can be expressed in the
form of a Green-Kubo formulae for the flux-flux auto correlation
function calculated at the transition state
\cite{hangg,keck,kap,chand,yam,tan,san} which acts as a dividing
surface between the reactant and the product states. The theory
of reactive flux has been developed both in classical and quantum
mechanical contexts to obtain the rate coefficient of a chemical
reaction. Recently Kohen and Tannor \cite{tan} have developed a
method to calculate the classical rate coefficient and hence the
transmission coefficient as a function of time in terms of phase
space distribution function in the high friction regime. The
method is simple and directly addresses the relevant
nonequilibrium dynamics at the barrier top. The theory
\cite{tan,san} not only recovers the Kramers-Grote-Hynes
transmission coefficient in the long time or asymptotic limit but
also traces the nature of its transient dynamics and its approach
towards the steady state.

The key element of the Kohen-Tannor formulation \cite{tan} is
that it allows us to express the correlation function with the
help of a probability distribution function which describes
dynamics of the reaction co-ordinate undergoing Brownian motion
in phase space at the barrier top without resorting to a solution
of the generalized Fokker-Planck equation. It is worthwhile in
this context to ask whether the theory can be extended to quantum
domain to include the generic quantum mechanical effects. The
essential requirement for such a theory is obviously the quantum
phase space distribution function. Very recently, based on a
coherent state representation of quantum noise operator and an
ensemble averaging procedure, we have developed
\cite{db1,skb,db2,db3} a c-number quantum generalized Langevin
equation. The theory is valid for arbitrary temperature and noise
correlation. We extend the scheme in the present context to make
use of a true quantum probability distribution function of the
relevant dynamical variables in c-numbers and express the rate
constant or transmission coefficient accordingly. Our object is
thus two-fold, i.e. ,

(i) to express the quantum mechanical rate constant in terms of
the c-number quantum phase space functions.

(ii) to derive a time dependent quantum Kramers-Grote-Hynes (KGH)
transmission coefficient which is valid for arbitrary noise
correlation and temperature for intermediate to strong friction
regime.

Two pertinent points are now in order. First, generally speaking
the phase space distribution functions used in quantum mechanical
contexts, e.g. , in quantum optics are quasi-probability
distribution functions like Wigner, or Glauber-Sudarshan
functions \cite{loui}. They often become singular or negative in
the fully quantum domain and can therefore not be regarded as true
probability distribution functions as their classical
counterparts. We stress that the present formalism is free from
such difficulties. Second, the expression for the rate constant
as a time integral over flux-flux auto correlation function has
been evaluated earlier both analytically and numerically using
path integral approaches. Although for equilibrium properties the
imaginary time propagator has proved to be very much successful
\cite{bern}, particularly for developing quantum transition state
theory \cite{yam}, it is not easy to extract dynamical
information for nonequilibrium problems \cite{topa} because of the
oscillatory nature of the real time propagator in many
situations. The c-number phase space method that we use here, on
the other hand, being independent of path integral approach is an
alternative for calculation of the rate without these problems.

The outlay of the paper is as follows; After a brief review of our
recent scheme for quantum Brownian motion we introduce the
conditional probability function of c-numbers at the barrier top
and an equilibrium distribution function at the reactant well in
Sec.II. In the following section the reactive flux expression for
rate constant is formulated in terms of these c-number phase
space functions. The time dependent general quantum
Kramers-Grote-Hynes transmission coefficient has been derived in
the spirit of Kohen and Tannor formalism. We work out in detail
the Markovian and non-Markovian limits of this expression. The
paper is concluded in Sec.IV.

\section{Quantum Brownian motion ; Evolution of c-number
phase space functions}

\subsection{Quantum Langevin equation in c-numbers}

The kinetics of a particle undergoing a chemical reaction in a
medium can be described by the following generic system-reservoir
Hamiltonian

\begin{equation}\label{1}
H = \frac{\hat{p}^2}{2} + V(\hat{q}) +
\sum_j\left\{{\frac{\hat{p}_j^2}{2} +
\frac{1}{2}\kappa_j(\hat{q}_j - \hat{q})^2}\right\}
\end{equation}
Here $\hat{q}$ and $\hat{p}$ are the co-ordinate (here considered
as reaction co-ordinate) and momentum operators of the particle,
respectively and the set $\{\hat{q}_j,\hat{p}_j\}$ is the set of
co-ordinate and momentum operators for the reservoir oscillators
linearly coupled to the system through coupling constant
$\kappa_j$. $V(\hat{q})$ denotes the external potential field
which, in general, is nonlinear. The Hamiltonian contains
counter-terms quadratic in $\hat{q}$ to ensure the independence
of barrier height and coupling constants. The operators follow
the usual commutation relations $[\hat{q},\hat{p}] = i\hbar$ and
$[\hat{q}_i,\hat{p}_j] = i\hbar\delta_{ij}$. Over the last
several decades the model and its variants have widely served as
the fundamental paradigm for quantum Brownian motion in chemical
dynamics, condensed matter physics, quantum optics and allied
areas within various approximate descriptions. Our primary aim in
this section is to search for a true c-number phase space
function which describes the evolution of quantum Brownian motion
of the particle at the barrier top and its thermalization in the
reactant well.

Eliminating the reservoir degrees of freedom in the usual way we
obtain the operator Langevin equation for the particle

\begin{equation}\label{2}
\ddot{ \hat{q} } + \int_0^t dt' \gamma(t-t') \dot{ \hat{q} } (t')
+ V' (\hat{q}) = \hat{F} (t)
\end{equation}

where the noise operator $\hat{F} (t)$ and the memory kernel
$\gamma (t)$ are given by

\begin{equation}\label{3}
\hat{F} (t) = \sum_j \left[ \{ \hat{q}_j (0) - \hat{q} (0)\}
\kappa_j \cos\omega_jt + \hat{p}_j (0) \kappa_j^{1/2}
\sin\omega_jt \right]
\end{equation}
and
\begin{eqnarray*}
\gamma (t) = \sum_j \kappa_j \cos\omega_jt\nonumber
\end{eqnarray*}
or in the continuum limit
\begin{equation}\label{4}
\gamma(t) = \int_0^\infty \kappa(\omega) \rho(\omega) \cos\omega
t \; d\omega
\end{equation}

with $\kappa_j = \omega_j^2$. The masses have been assumed to be
unity. $\rho (\omega)$ represents the density of the reservoir
modes.

Eq.(\ref{2}) is an exact Langevin equation for which the noise
properties of $\hat{F}(t)$ can be defined using a suitable
canonical initial distribution of bath co-ordinates and momentum.
Our first task is to replace it by an equivalent generalized
quantum Langevin equation (GQLE) in c-numbers
\cite{db1,skb,db2,db3}. To achieve this we proceed in two steps.
We first carry out the {\it quantum mechanical average} of
Eq.(\ref{2}).

\begin{equation}\label{5}
\langle \ddot{ \hat{q} } \rangle + \int_0^t \;dt' \gamma(t-t')
\langle \dot{ \hat{q} }(t') \rangle + \langle V'( \hat{q} )
\rangle = \langle \hat{F} (t) \rangle
\end{equation}

Where the average $\langle....\rangle$ is taken over the initial
product separable quantum states of the particle and the bath
oscillators at $t=0$, $|\phi \rangle \{ |\alpha_1\rangle
|\alpha_2\rangle.......|\alpha_N\rangle \}$. Here $|\phi\rangle$
denotes any arbitrary initial state of the particle and
$|\alpha_i\rangle$ corresponds to the initial coherent state of
the i-th bath oscillator. $|\alpha_i\rangle$ is given by
$|\alpha_i \rangle = \exp(-|\alpha_i|^2/2) \sum_{n_i=0}^\infty
(\alpha_i^{n_i} /\sqrt{n_i !} ) | n_i \rangle $, $\alpha_i$ being
expressed in terms of the mean values of the co-ordinate and
momentum of the i-th oscillator, $\langle \hat{q}_i (0) \rangle =
( \sqrt{\hbar} /2\omega_i) (\alpha_i + \alpha_i^\star )$ and
$\langle \hat{p}_i (0) \rangle = i \sqrt{\hbar\omega_i/2 }\;
(\alpha_i^\star - \alpha_i )$, respectively. It is important to
note that $\langle \hat{F} (t) \rangle$ is a classical-like noise
term which, in general, is a nonzero number because of the quantum
mechanical averaging over the co-ordinate and momentum operators
of the bath oscillators with respect to initial coherent states
and arbitrary initial state of the particle and is given by

\begin{equation}\label{6}
\langle \hat{F} (t) \rangle = \sum_j \left[ \{ \langle
\hat{q}_j(0)\rangle - \langle \hat{q}(0)\rangle \} \kappa_j \cos
\omega_j t + \langle \hat{p}_j(0) \rangle \kappa_j^{1/2} \sin
\omega_j t \right]
\end{equation}

It is convenient to write Eq.(\ref{5}) as follows;

\begin{equation}\label{7}
\langle \ddot{ \hat{q} } \rangle + \int_0^t dt'\; \gamma(t-t')
\langle \dot{\hat{q}}(t') \rangle + \langle V'( \hat{q} ) \rangle
= f(t)
\end{equation}

where we let the quantum mechanical mean value $\langle \hat{F}
(t) \rangle = f(t)$. We now turn to the ensemble averaging. To
realize $f(t)$ as an effective c-number noise we now assume that
the momentum $\langle \hat{p}_j(0) \rangle$ and the co-ordinate
$\langle \hat{q}_j(0)\rangle - \langle \hat{q}(0)\rangle$ of the
bath oscillators are distributed according to a canonical
distribution of Gaussian form as,

\begin{equation}\label{8}
{\cal P}_j = {\cal N} \exp \left \{ - \frac{  [ \langle \hat{p}_j
(0) \rangle^2 + \kappa_j \left \{ \langle \hat{q}_j (0) \rangle -
\langle \hat{q} (0) \rangle \right \}^2 ] }{ 2 \hbar \omega_j
\left ( \bar{n}_j + \frac{1}{2} \right ) } \right \}
\end{equation}

so that for any quantum mean value ${\cal O}_j ( \langle\hat{p}_j
(0) \rangle, \{ \langle \hat{q}_j (0) \rangle  - \langle \hat{q}
(0) \rangle \} )$, the statistical average $\langle....\rangle_s$
is

\begin{equation}\label{9}
\langle {\cal O}_j \rangle_s = \int {\cal O}_j\; {\cal P}_j\;
d\langle \hat {p}_j (0) \rangle \; d\{ \langle \hat{q}_j(0)
\rangle - \langle \hat{q} (0) \rangle \}
\end{equation}

Here $\bar{n}_j$ indicates the average thermal photon number of
the j-th oscillator at temperature $T$ and is given by
Bose-Einstein distribution $\bar{n}_j=1 / [\exp (\hbar
\omega_j/kT) - 1]$ and ${\cal N} $ is the normalization constant.

The distribution Eq.(\ref{8}) and the definition of the
statistical average over quantum mechanical mean values
Eq.(\ref{9}) imply that $f(t)$ must satisfy

\begin{equation}\label{10}
\langle f (t) \rangle_s = 0
\end{equation}

and

\begin{eqnarray*}
\langle f(t) f(t') \rangle_s = \frac {1} {2} \sum_j \kappa_j \;
\hbar \omega_j  \left ( \coth \frac {\hbar \omega_j } {2 k T}
\right ) \cos \omega_j (t-t')
\end{eqnarray*}

or in the continuum limit

\begin{eqnarray}\label{11}
\langle f(t) f(t') \rangle_s & = & \frac {1} {2} \int_0^\infty
d\omega \;\kappa(\omega)\; \rho(\omega) \;\hbar \omega \left (
\coth \frac {\hbar \omega } {2 k T} \right ) \cos \omega
(t-t')\nonumber\\ & \equiv & c ( t- t' )
\end{eqnarray}

That is, c-number noise $f(t)$ is such that it is zero-centered
and satisfies standard fluctuation-dissipation relation as known
in the literature.

We now add the force term $V' ( \langle \hat q \rangle )$ on both
sides of the Eq.(\ref{7}) and rearrange it to obtain

\begin{equation}\label{12}
\ddot{q} (t)  + V'(q) + \int_0^t  dt' \; \gamma (t-t')\;
\dot{q}(t')  = f(t) + Q ( t )
\end{equation}

where we put $\langle \hat q (t) \rangle = q (t)$ and $\langle
\dot{ \hat q }(t) \rangle = p (t)$ ; $q(t)$ and $p (t)$ being
quantum mechanical mean values and also

\begin{equation}\label{13}
Q (t) = V' (q) -\langle V' ( \hat {q} ) \rangle
\end{equation}

represents the quantum correction to classical potential.

Eq.(\ref{12}) offers a simple interpretation. This is that GQLE
Eq.(\ref{12}) is governed by a c-number quantum noise $f( t )$
due to the heat bath, characterized by the properties
Eq.(\ref{10}), Eq.(\ref{11}) and a quantum correction term $Q ( t
)$ characteristic of nonlinearity of the potential. Our general
results contain $Q(t)$ in all orders formally. However for a
practical calculation we need a recipe for calculation of $Q ( t
)$. This has been discussed earlier in several contexts
\cite{db1,skb,db2,db3,sm,akp}. For the present purpose we
summarize it as follows:

Referring to the quantum mechanics of the system in the Heisenberg
picture one may write,

\begin{eqnarray}\label{14}
\hat{q} (t) & = & \langle\hat{q} (t) \rangle + \delta\hat{q}
(t)\nonumber\\
\hat{p}(t) & = & \langle\hat{p} (t) \rangle + \delta\hat{p} (t)
\end{eqnarray}

$\delta\hat{q} (t)$ and $\delta\hat{p} (t)$ are the operators
signifying quantum corrections around the corresponding quantum
mechanical mean values $q$ and $p$. By construction $\langle
\delta\hat{q} \rangle = \langle \delta\hat{p} \rangle = 0$ and $[
\delta\hat{q},\delta\hat{p} ] = i\hbar$. Using Eq.(\ref{14}) in
$\langle V' ( \hat {q} ) \rangle$ and a Taylor series expansion
around $\langle\hat{q} \rangle$ it is possible to express $Q (t)$
as

\begin{equation}\label{15}
Q(t) = -\sum_{n \ge 2} \frac{1}{n!} V^{(n+1)} (q)
\langle\delta\hat{q}^n(t)\rangle
\end{equation}

Here $V^{(n)} (q)$ is the n-th derivative of the potential $V (
q)$. To second order $Q (t)$ is given by $Q(t) = -\frac{1}{2}
V^{\prime\prime\prime} (q) \langle \delta \hat{q}^2(t) \rangle$
where $q(t)$ and $\langle \delta \hat{q}^2 (t) \rangle$ can be
obtained as explicit functions of time by solving following set
of approximate coupled equations Eq.(\ref{16}) to Eq.(\ref{18})
together with Eq.(\ref{12})

\begin{eqnarray}
\dot{\langle \delta \hat{q}^2 \rangle} & = & \langle \delta
\hat{q} \delta \hat{p} + \delta \hat{p} \delta \hat{q} \rangle\label{16}\\
\dot{\langle \delta \hat{q} \delta \hat{p} + \delta \hat{p}
\delta \hat{q} \rangle} & = & 2\langle \delta \hat{p}^2 \rangle -
2V^{\prime\prime}(q) \langle \delta \hat{q}^2 \rangle\label{17}\\
\dot{\langle \delta \hat{p}^2 \rangle} & = & -V^{\prime\prime}(q)
\langle \delta \hat{q} \delta \hat{p} + \delta \hat{p} \delta
\hat{q} \rangle\label{18}
\end{eqnarray}

While the above set of equations provide analytic solutions
containing lowest order quantum corrections, the successive
improvement of $Q(t)$ can be achieved by incorporating higher
order contribution due to potential $V(q)$ and the higher order
effect of dissipation on the quantum correction terms. In Appendix
A we have derived the equations for quantum corrections upto
forth oder \cite{sm}. Under very special circumstances, it has
been possible to include quantum effects to all orders
\cite{db3,akp}. The present theory thus takes into account of the
anharmonicity as an integral part of the treatment. This is
somewhat different from usual classical theories where
anharmonicity is incorporated from outside as a finite barrier
corrections \cite{talk,melni}. We mention here that centroid
molecular dynamics simulation methods have been developed to
calculate the expression for rate constant which is related to
flux-flux correlation functions involving two nonlinear operators
within bilinear system-reservoir coupling \cite{voth}. Apart from
anharmonicity of the system potential it is also possible to
consider nonlinear coupling, e.g. , $\sum_{j} \frac{1}{2}\;
\kappa_{j}\; f(\hat{q}) \;\hat{q}_{j}$ (beyond bilinear) between
the system and the reservoir. In the present context we, however,
envisage two difficulties. First, the potential in which the
particle moves gets significantly modified because of the absence
of any counter term in the Hamiltonian as in Eq.(\ref{1}).
Secondly, the operator Langevin equation of motion would contain
multiplicative noise for which a linear dissipation fails to
satisfy the usual fluctuation-dissipation relation. Because of
these reasons we confine ourselves within traditional bilinear
coupling. The method can, however, be readily extended to
anharmonic bath comprising, say, of Morse oscillators or of two
level systems, for which coherent states are well known (Morse
oscillator is amenable to a theoretical description within SU(2)
Lie algebra and the associated coherent states \cite{dsr}, while
the two-level system in terms of Pauli matrix and Radcliffe
coherent states \cite{rad}) for constructing quantum Langevin
equation since the quantum mechanical averaging with these states
can be computed exactly in the same way as we proceeded from
Eq.(\ref{2}) to obtain Eq.(\ref{5}). The present approach has
recently been utilized by us to derive \cite{db1,db2} generalized
quantum Kramers' equation for calculation of escape rate over the
barrier by extending the flux-over-population method to quantum
domain.

\subsection{Conditional probability function of c-numbers at the
barrier top}

We now show that the present scheme allows us to derive a quantum
conditional probability function in c-numbers $W(q, p, t; q_0,
p_0, t=0)$ for an ensemble of trajectories starting from a
specific initial condition $q=q_0$, $p=p_0$ at $t=0$ and
following the quantum Langevin dynamics Eq.(\ref{12}). Since we
will be primarily concerned with the reactive flux at the barrier
top located at $q=q_0(=0)$ it is convenient to linearize the
potential $V(q)$ at this point as $V(q)=V(0)-\frac{1} {2}
\omega_b^2q^2$ where $\omega_b^2=V''|_{q_0=0}$ refers to the
oscillator frequency of the inverted well at $q=q_0$. We are then
lead to the following Langevin equation;

\begin{equation}\label{19}
\ddot{q} -\omega_b^2 q   + \int_0^t  dt' \; \gamma (t-t')\;
\dot{q}(t') = f(t) + Q ( t )
\end{equation}

Laplace transform of Eq.(\ref{19}) yields the formal solutions for
$q(t)$ and $p(t)$ as follows;

\begin{equation}\label{20}
q (t) = \langle q (t) \rangle_s + \int_0^t C_v(t-\tau) f(\tau)\;
d\tau
\end{equation}

\begin{equation}\label{21}
p (t) = \langle p (t) \rangle_s + \int_0^t \dot{C}_v(t-\tau)
f(\tau)\; d\tau
\end{equation}

where

\begin{equation}\label{22}
\langle q (t) \rangle_s = p_0\; C_v(t) + q_0\; C_q(t) + G(t)
\end{equation}

\begin{equation}\label{23}
\langle p (t) \rangle_s = p_0\; \dot{C}_v(t) + q_0\; \dot{C}_q(t)
+ \dot{G}(t)
\end{equation}

\begin{equation}\label{24}
G (t) = \int_0^t C_v(t-\tau)\; Q(\tau)\; d\tau
\end{equation}

\begin{equation}\label{25}
C_q (t) = 1 + \omega_b^2 \int_0^t C_v(\tau)\;d\tau
\end{equation}

Here $C_v(t)$ is the inverse Laplace transform of
$\tilde{C}_v(\mu)$, i.e.

\begin{equation}\label{26}
C_v(t) = L^{-1} \left[\mu^2+\mu \;\tilde{\gamma} (\mu) -
\omega_b^2\right]^{-1}
\end{equation}

with

\begin{equation}\label{27}
\tilde{\gamma} (\mu) = \int_0^t dt\; \gamma(t)\; e^{-\mu t}
\end{equation}

the Laplace transform of $\gamma(t)$. Several remarks are now in
order.

First, the solutions of Eq.(\ref{20}) and Eq.(\ref{21}) of the
Langevin equation Eq.(\ref{12}) are the quantum mechanical mean
values distributed around their statistical average  $\langle q
\rangle_s$ and $\langle p \rangle_s$. The distribution is due to
Gaussian, c-number quantum noise $f(t)$ which obeys Eq.(\ref{10})
and Eq.(\ref{11}).

Second, the averages $\langle q \rangle_s$ and $\langle p
\rangle_s$ not only depend on the relaxation functions $C_v(t)$
and $C_q(t)$ in the same way as in the corresponding classical
theory \cite{tan} but also on the convolution integral $G(t)$
which is a quantum contribution and takes care of the interplay of
dissipation and nonlinearity in the dynamics. It is important to
note that although we have linearized the potential $V(q)$,
$Q(t)$ remains finite and nonzero. This is because linearization
involves local dynamics around a specific $q$ (say, around $q=0$
for the barrier top) which is a quantum mechanical mean value
$\langle \hat{q} \rangle$. $Q(t)$ as expressed in Eq.(\ref{15})
implies quantum fluctuation around this mean value manifested as
shifts in $\langle q \rangle_s$ and $\langle p \rangle_s$ as
shown in Eq.(\ref{22}) and Eq.(\ref{23}) respectively. Making the
quantum fluctuations $\langle\delta\hat{q}^n(t)\rangle$ or $Q(t)$
equal to zero would mean that the quantum mechanical position of
the particle is exactly specified. Thus nonzero finite quantum
fluctuation around $q=0$ takes care of uncertainty relation. In
other words to consider the conditional probability distribution
function which originates as a $\delta$-function in position
(means quantum mechanical mean) at the barrier top, we must have
to associate a finite minimum quantum fluctuation or uncertainty
around this point.

Third, since no approximation has been made on $Q(t)$, the
solutions Eq.(\ref{20}) and Eq.(\ref{21}) for $q(t)$ and $p(t)$
formally take care of quantum effects to all orders. In actual
practice, however, they can be calculated order by order to a high
degree of accuracy.

Fourth, the quantum effects enter into the dynamics in two
different ways. The quantum nature of the system is manifested
through nonlinearity of the potential while the heat bath imparts
quantum noise in a thermal environment.

$f(t)$ is a c-number Gaussian random process. Since the sum of
the random processes is Gaussian, so is the integral in
Eq.(\ref{20}) and Eq.(\ref{21}) we conclude that $q-\langle q
\rangle_s$ and $p-\langle p \rangle_s$ are Gaussian random
processes. Therefore the first and the second moments suffice to
determine the distribution function. Making use of Eq.(\ref{20})
and Eq.(\ref{21}) we construct the second moments of $q - \langle
q \rangle_s \left( = \int_0^t C_v(t-\tau) f(\tau)\; d\tau \right)$
and $p  - \langle p  \rangle_s \left( = \int_0^t
\dot{C}_v(t-\tau) f(\tau)\; d\tau \right)$ as

\begin{equation}\label{28}
A_{11}(t) = \langle (q - \langle q  \rangle_s)^2 \rangle_s = 2
\int_0^t C_v(t_1)\; dt_1 \int_0^{t_1} C_v(t_2) \;c(t_1-t_2)\; dt_2
\end{equation}

\begin{equation}\label{29}
A_{22}(t) = \langle (p - \langle p  \rangle_s)^2 \rangle_s = 2
\int_0^t \dot{C}_v(t_1)\; dt_1 \int_0^{t_1} \dot{C}_v(t_2)
\;c(t_1-t_2)\; dt_2
\end{equation}

\begin{equation}\label{30}
A_{12}(t) = A_{21}(t) = \langle \;(q - \langle q  \rangle_s) (p -
\langle p  \rangle_s)\; \rangle_s = 2 \int_0^t C_v(t_1)\; dt_1
\int_0^{t_1} \dot{C}_v(t_2)\; c(t_1-t_2)\; dt_2
\end{equation}

We are now in a position to write down the probability
distribution function of $q$, $p$ conditional on $q(t=0)=q_0$ and
$p(t=0)=p_0$

\begin{eqnarray}{\label{31}}
W(q, p, t; q_0, p_0, t=0) & = & \frac{1} {2 \pi (det A)^{1/2} }
\times \exp \left[-\; \frac{1}{2(det A)} \left\{ A_{22} (q -
\langle q \rangle_s)^2  \right. \right. \nonumber \\
& - & \left. \left. 2 A_{12} (q - \langle q  \rangle_s) (p -
\langle p  \rangle_s) + A_{11} (p - \langle p  \rangle_s)^2
\right\}  \right]
\end{eqnarray}

Here $detA$ refers to the determinant of the matrix $A$. The
conditional probability $W$ therefore describes the transition
probability in a c-number quantum phase space. Two remarks are
pertinent at this point. First, it is easy to check that the $W$
reduces to the classical distribution function in the limit
$\hbar\rightarrow0$, i.e. , when $G(t)\rightarrow0$ and
$\frac{\hbar \omega_b} {2} \coth ( {\hbar \omega_b}/{2 k T})$ goes
over to $kT$. Second, since $W(q, p, t; q_0, p_0, t=0)$ concerns
dynamics at the barrier top where strong nonequilibrium situation
prevails, $W$ does not go over to an equilibrium distribution
because of hyperbolic nature of the trajectories emanating from
$q=q_0=0$. However, this phase space distribution function serves
as a basis for calculation of average of the relevant dynamical
quantities at the barrier top.

\subsection{Equilibrium distribution in the reactant well }

In addition to the distribution at the barrier top $W(q, p, t;
q_0, p_0, t=0)$ it is necessary to consider further the
equilibrium distribution of the particles in the reactant well.
As usual a good description of the kinetics can be obtained by
linearizing the potential $V(q)$ at the bottom of the well (say
at $q=0$ ) so that $V(q)=V(0)+\frac{1} {2} \omega_0^2q^2$ where
$\omega_0$ corresponds the frequency at the bottom of the
reactant well. The Langevin dynamics can be obtained from
Eq.(\ref{19}) with the replacement of $-\omega_b^2$ by
$\omega_0^2$. The essential changes that are needed for the
solution of $q(t)$ and $p(t)$ are

\begin{equation}\label{32}
C_v(t) = L^{-1} \tilde{C}(\mu) = L^{-1} \left[\mu^2+\mu\;
\tilde{\gamma} (\mu) + \omega_0^2\right]^{-1}
\end{equation}

\begin{equation}\label{33}
C_q (t) = 1 - \omega_0^2 \int_0^t C_v(\tau)\; d\tau
\end{equation}

The first and second moments of the distribution $W$ can
therefore be calculated by employing the relations Eq.(\ref{20})
to Eq.(\ref{24}) and Eq.(\ref{28}) to Eq.(\ref{30}). Since $W(q,
p, t; q_0, p_0, t=0)$ pertains to the dynamics in the reactant
well where the trajectories evolve around a stable elliptic fixed
point, the distribution $W$ eventually reaches an equilibrium
thermal distribution in the long time limit, i.e. ,

\begin{equation}\label{34}
{\cal L}t_{t\rightarrow\infty} W(q, p, t; q_0, p_0, t=0) =
W_{eqm} (q, p)
\end{equation}

Since an equilibrium distribution does not depend on the way in
which the final state is reached, one may consider the Markovian
limit for simplicity but without any loss of generality. In this
limit the density distribution of the reservoir modes is flat,
i.e. , $\kappa(\omega) \rho(\omega) = {2\gamma}/\pi$, $\gamma$
being dissipation constant. This results in

\begin{equation}\label{35}
\gamma (t) =2\; \gamma \;\delta (t)
\end{equation}

and

\begin{equation}\label{36}
c(t) = \frac{1} {2}\; \gamma \;\hbar \omega_0 \;\coth \left(
\frac{\hbar \omega_0}{2 kT}\right)\; \delta(t)
\end{equation}

from Eq.(\ref{4}) and Eq.(\ref{11}), respectively. From
Eq.(\ref{35}) we have $\tilde{\gamma}(\mu)=\gamma$ which when put
in Eq.(\ref{32}) gives after inverse Laplace transform

\begin{equation}\label{37}
C_v (t) = \frac{\left(e^{\mu_1^\prime t} - e^{\mu_2^\prime
t}\right)}{{2 \omega_1^\prime}}
\end{equation}

where $\mu_1^\prime = - \frac{\gamma} {2} + \omega_1^\prime$ and
$\mu_2^\prime = - \frac{\gamma} {2} - \omega_1^\prime$ with
$\omega_1^\prime = \sqrt{\frac{\gamma^2} {4} - \omega_0^2}$. By
Eq.(\ref{33}) we obtain

\begin{equation}\label{38}
C_q (t) = 1 - \frac{\omega_0^2}{2 \omega_1^\prime} \left[\frac{1}
{\mu_1^\prime}(e^{\mu_1^\prime t} - 1) - \frac{1}
{\mu_2^\prime}(e^{\mu_2^\prime t} - 1) \right]
\end{equation}

It is easy to check the stationary limits for the model

\begin{eqnarray}
{\cal L}t_{t\rightarrow\infty} C_v (t) & = & 0 \;\;\;\;\;\;\;\;\;
{\cal L}t_{t\rightarrow\infty} C_q (t)  =  0\label{39}\\
{\cal L}t_{t\rightarrow\infty} \dot{C}_v (t) & = & 0
\;\;\;\;\;\;\;\;\; {\cal L}t_{t\rightarrow\infty} \dot{C}_q (t) =
0\label{40}
\end{eqnarray}

An important quantity that contains quantum contribution due to
the system is $G(t)$ as defined in Eq.(\ref{24}). Putting
Eq.(\ref{37}) in Eq.(\ref{24}) we write

\begin{equation}\label{41}
G(t) = \frac{1} {2 \omega_1^\prime} \left[ \int_0^t
e^{\mu_1^\prime \tau} Q (t - \tau)\; d\tau - \int_0^t
e^{\mu_2^\prime \tau} Q (t - \tau)\;d\tau\right]
\end{equation}

For the bound system, i.e. , in the neighborhood of the minimum
of the well the solutions $q(t)$ and $\langle \delta \hat{q}^n
(t) \rangle$, in general, are oscillating in nature, while the
exponentials in Eq.(\ref{41}) are decaying in time. Keeping in
mind the nature of the system, and expanding $Q(t-\tau)$ in
Taylor series as $Q(t-\tau) = \sum_m \frac{(-1)^m} {m!}\; Q^m(t)
\;\tau^m$, $Q^m$ being the m-th time derivative of $Q(t)$, the
long time limit of $G(t)$ is given by

\begin{equation}\label{42}
{\cal L}t_{t\rightarrow\infty} G(t) = \frac{1} {2
\omega_1^\prime}\left[ \sum_m (-1)^m Q^m(t) \left(\frac{1}
{(-\mu_1^\prime)^m} - \frac{1} {(-\mu_2^\prime)^m}\right)\right]
\end{equation}

Since from Eq.(\ref{15}) the leading order quantum correction is
given by $Q(t)=-\frac{1} {2} V'''(q) \langle \delta \hat{q} (t)^2
\rangle$ it is now easy to check that for the time derivative
terms in Eq.(\ref{42}), this correction leads to (and also for
higher order correction terms in $Q(t)$),

\begin{equation}\label{43}
{\cal L}t_{t\rightarrow\infty} G(t) = 0
\end{equation}

for quartic nonlinearity, where we have made use of the relations
Eq.(\ref{22}) and Eq.(\ref{23}). Similarly one may show

\begin{equation}\label{44}
{\cal L}t_{t\rightarrow\infty} \dot{G}(t) = 0
\end{equation}

By virtue of the relations Eq.(\ref{39}), Eq.(\ref{40}),
Eq.(\ref{43}) and Eq.(\ref{44}) we have

\begin{equation}\label{45}
\langle q \rangle_s = 0 \;\;\;\;\;\;\langle p \rangle_s = 0
\end{equation}

Furthermore making use of the relation Eq.(\ref{36}) for $c(t)$ we
obtain from Eq.(\ref{28}) to Eq.(\ref{30}) the following
expressions for the second moments.

\begin{eqnarray}
A_{11} (t) & = & \gamma \;\hbar \omega_0 \coth \left( \frac{\hbar
\omega_0}{2 k T}\right)\int_0^t C_v^2(t_1)\; dt_1\label{46}\\
A_{22} (t) & = & \gamma \;\hbar \omega_0 \coth \left( \frac{\hbar
\omega_0}{2 k T}\right)\int_0^t \dot{C}_v^2(t_1)\; dt_1\label{47}\\
A_{12}(t) = A_{21} (t)& = & \gamma \;\hbar \omega_0 \coth \left(
\frac{\hbar \omega_0}{2 k T}\right)\int_0^t C_v(t_1)
\;\dot{C}_v(t_1)\; dt_1\label{48}
\end{eqnarray}

Putting Eq.(\ref{37}) in Eq.(\ref{46}) to Eq.(\ref{48}) we obtain
the long time limits of the second moments as

\begin{eqnarray}
{\cal L}t_{t\rightarrow\infty} A_{11} (t) & = & \frac{ \hbar} {2
\omega_0} \coth \left( \frac{\hbar \omega_0}{2 k T}\right)\label{49}\\
{\cal L}t_{t\rightarrow\infty} A_{22} (t) & = & \frac{ \hbar
\omega_0}{2} \coth \left( \frac{\hbar \omega_0}{2 k T}\right)\label{50}\\
{\cal L}t_{t\rightarrow\infty} A_{12} (t) & = & {\cal
L}t_{t\rightarrow\infty} A_{21} (t) = 0\label{51}
\end{eqnarray}

Next we use the relations Eq.(\ref{45}) and Eq.(\ref{49}) to
Eq.(\ref{51}) in Eq.(\ref{31}) to obtain the equilibrium
distribution of quantum mechanical mean values, $q$, $p$ as

\begin{equation}\label{52}
W_{eqm} (q, p) = Z^{-1} \exp \left[-\frac{\frac{p^2}{2} +
V(q)}{\frac{ \hbar \omega_0}{2} \coth \left( \frac{\hbar
\omega_0}{2 k T}\right)}\right]
\end{equation}

where $Z^{-1}$ is the normalization constant and $V(q)$ has the
linearized form in the neighborhood of the bottom of the reactant
well characterized by the frequency $\omega_0$. The width of the
distribution $\frac{ \hbar \omega_0}{2} \coth \left( \frac{\hbar
\omega_0}{2 k T}\right)$ reduces to $k T$ in the classical limit
and one obtains the usual Maxwell-Boltzmann distribution. We thus
note that Eq.(\ref{52}) has the same form as Eq.(\ref{8}) (since,
$\overline{n}(x)+1/2 = \frac{1}{2}\coth(x/2)$). In fact since theq
potential at the reactant well is assumed to the harmonic, it
seems consistent to use the same ansatz that was used for the
harmonic bath modes. This has turned out to be the case.

The c-number conditional probability distribution $W(q, p, t;
q_0, p_0, t=0)$ describing the dynamics at the barrier top and
the thermal distribution $W_{eqm}(q, p)$ describing equilibrium
in the reactant well are the two main results of this section.
They are primary elements for calculation of statistical averages
used in the expression for reactive flux that follows in the next
section.

\section{c-number phase space function formulation of
reactive flux theory}

\subsection{A general expression for quantum transmission coefficient}

We now consider $q$ as the reaction co-ordinate and the potential
$V(q)$ defines the dividing surface conveniently located at the
transition state $q=q_0$ which separates the reactant state from
the product state. While the reactant particles are thermally
equilibriated deep inside the reactant well strong nonequilibrium
situation prevails at $q=q_0$. The key element in the reactive
flux formulation is that the nonequilibrium concentration relaxes
to the equilibrium concentration on a timescale at which the
equilibrium correlation function decays. Based on a timescale $t$
obeying $\frac{1} {\Gamma} \gg t \gg \tau_c$ where $\frac{1}
{\Gamma}$ is the decay time of equilibrium fluctuation and
$\tau_c$ is correlation time, the reactive flux expression for
rate coefficient in the form of Green-Kubo transport coefficient
has been derived by various workers
\cite{keck,kap,chand,yam,tan,san}. Our aim in this section is to
provide a {\it c-number quantum phase space} representation of
the rate coefficient. The present formulation is a simple quantum
translation of the reactive flux formulation advocated by Kohen
and Tannor \cite{tan} a few years ago.

In the spirit of classical reactive flux theory the quantum rate
coefficient $k(t)$ can be expressed in terms of the following
correlation function

\begin{eqnarray}
k (t) & = & \frac{\langle \dot{\theta}_P (q_0)\; \theta_P [ q
(q_0, p_0, t)] \rangle_{qs}}{\langle \theta_R (q_0) \rangle_{qs}}\label{53}\\
& = & \frac{\langle p_0 \;\delta(q_0)\; \theta_P [ q (q_0, p_0,
t)] \rangle_{qs}}{\langle \theta_R (q_0) \rangle_{qs}}\label{54}
\end{eqnarray}

Here $q$ is the quantum mechanical mean values of reaction
co-ordinate and $p$ is the corresponding momentum. $\theta_P[q]$
is a step function implying $1$ if $q>0$ and $0$ otherwise.
Noting that $R$ and $P$ refer to the reactant and transition
states, respectively, we have $\theta_R + \theta_P=1$. $q_0$,
$p_0$ correspond to initial co-ordinate (at the barrier top;
$q=q_0=0$) and momentum, respectively. The averaging $\langle ...
\rangle_{qs}$ is carried over an initial equilibrium
distribution, which in our case is the c-number quantum
distribution $W_{eqm}(q, p)$ of $q$, $p$ as given by
Eq.(\ref{52}). The subscript 'qs' in the average is labeled to
make a distinction from its classical counterpart. Thus when
written explicitly in terms of Eq.(\ref{52}), Eq.(\ref{54})
reduces to the following form

\begin{equation}\label{55}
k(t)=\frac{Z^{-1}}{\langle\theta_R
(q_0)\rangle}_{qs}\int_{-\infty}^{+\infty} dq_0
\int_{-\infty}^{+\infty} dp_0 \;\exp\left[-\frac{\frac{p_0^2}{2} +
V(q_0)} {\frac{ \hbar \omega_0}{2} \coth \left( \frac{\hbar
\omega_0}{2 k T}\right)}\right]p_0\;\delta(q_0)\;\theta_P [ q
(q_0, p_0, t)]
\end{equation}

For a single realization of a stochastic trajectory the
characteristic function $\theta_P [ q (q_0, p_0, t)]$ is either
$0$ or $1$. However for an ensemble of trajectories starting from
$q_0$, $p_0$, $\theta_P[q]$ must be replaced by an average
$\overline{\theta_P[q]}$, which may assume any value between $0$
and $1$. In explicit form $\overline{\theta_P[q]}$ signifies an
integral over the c-number quantum conditional probability
distribution function $W(q, p, t; q_0, p_0, t=0)$ for the
ensemble of particles as given by Eq.(\ref{31}) so that

\begin{equation}\label{56}
\overline{\theta_P [ q (q_0, p_0,
t)]}=\chi(p_0,t)=\int_0^{+\infty}dq\int_{-\infty}^{+\infty}dp\;W(q,
p, t; q_0=0, p_0, t=0)
\end{equation}

$\chi$ is the usual "reactivity index" and takes care of
recrossing the transition state. With this replacement of
$\theta_P [ q (q_0, p_0, t)]$ by its average according to
Eq.(\ref{56}) we obtain the expression for quantum rate
coefficient as

\begin{eqnarray}
k(t) & = & \frac{Z^{-1}}{\langle\theta_R
(q_0)\rangle_{qs}}\int_{-\infty}^{+\infty} dq_0
\int_{-\infty}^{+\infty} dp_0\; \exp\left[-\frac{\frac{p_0^2}{2} +
V(q_0)} {\hbar \omega_0 (\bar{n}_0 + \frac{1} {2})}
\right]p_0\;\delta(q_0)\;\chi(q_0=0, p_0, t)\nonumber\\
& = & \frac{Z^{-1}}{\langle\theta_R
(q_0)\rangle_{qs}}\int_{-\infty}^{+\infty} dp_0\; p_0
\;\exp\left[-\frac{\frac{p_0^2}{2} + V(q_0=0)} {\hbar \omega_0
(\bar{n}_0 + \frac{1} {2})} \right]\chi(q_0=0, p_0, t)\label{57}
\end{eqnarray}

In order to extract out the dynamical contribution to the rate
coefficient it is convenient to define the latter as

\begin{equation}\label{58}
k(t) = \kappa(t)\; k_{QTST}
\end{equation}

where $\kappa(t)$ is the quantum transmission coefficient and
$k_{QTST}$ corresponds to the rate coefficient according to
transition state theory. In the following we outline an
expression for $k_{QTST}$.

It has been shown that thermalization of the reactant in the well
can be described by the distribution Eq.(\ref{52}). Assuming that
this equilibrium distribution holds good even at the barrier top
$q_0$ Eq.(\ref{52}) reduces to

\begin{equation}\label{59}
W_{eqm} (q_0=0, p)= Z^{-1} \exp\left[-\frac{\frac{p^2}{2} +
V(q_0=0)} {\hbar \omega_0 (\bar{n}_0 + \frac{1} {2})} \right]
\end{equation}

and we may calculate the rate coefficient as

\begin{equation}\label{60}
k_{QTST}=\int_0^\infty p\; W_{eqm} (q_0=0, p)\; dp
\end{equation}

which represents the positive average velocity of the particles
at the barrier top. The recrossing of the trajectories is
therefore completely ignored. Explicit integration in
Eq.(\ref{60}) using Eq.(\ref{59}) yields

\begin{equation}\label{61}
k_{QTST}=Z^{-1} \hbar \omega_0 (\bar{n}_0 + \frac{1} {2})
\exp\left[-\frac{V(q_0=0)} {\hbar \omega_0 (\bar{n}_0 + \frac{1}
{2})} \right]
\end{equation}

We now note that the total reactant concentration in the left
well is given by equilibrium average over the characteristic
function $\langle\theta_R (q_0)\rangle_{qs}$ which is obtained by

\begin{equation}\label{62}
\langle\theta_R (q_0)\rangle_{qs} = {\cal
L}t_{a\rightarrow\infty}\int_{q'-a}^{q'+a}dq
\int_{-\infty}^{+\infty} W_{eqm}(q, \; p)
\end{equation}

where $W_{eqm}(q, p)$ is given by Eq.(\ref{52}) with the
linearization of the potential well at $q=q'$ such that
$V(q)=V(q')+\frac{1} {2}\;\omega_0^2\;(q-q')^2$. By setting
$V(q')=0$ and calculating the normalization constant $Z^{-1}=2
\pi \hbar(\bar{n}_0 + \frac{1} {2})$, we have

\begin{equation}\label{63}
\langle\theta_R \rangle_{qs} = 1
\end{equation}

Now putting the value of $\langle\theta_R \rangle_{qs}$ in
Eq.(\ref{57}) and dividing $k(t)$ by $k_{QTST}$ as given in
Eq.(\ref{61}) we obtain the expression for quantum time dependent
transmission coefficient $\kappa(t)$ as;

\begin{equation}\label{64}
\kappa(t) = \frac{1} {\hbar \omega_0 (\bar{n}_0 + \frac{1} {2})}
\int_{-\infty}^{+\infty} p_0 \exp\left[-\frac{\frac{p_0^2}{2}}
{\hbar \omega_0 (\bar{n}_0 + \frac{1} {2})} \right]\chi(q_0=0,
p_0, t)\; dp_0
\end{equation}

where the fractional reactivity index $\chi$ is given by
Eq.(\ref{56}), i.e. ;

\begin{equation}\label{65}
\chi(p_0, t)=\int_0^{+\infty}dq\int_{-\infty}^{+\infty}\;dp\;W(q,
p, t; q_0=0, p_0, t=0)
\end{equation}

On explicit integration over $p$ using Eq.(\ref{31}) the above
expression gives

\begin{equation}\label{66}
\chi(p_0, t) = \frac{1} {(2\pi A_{11})^{1/2}} \int_0^\infty dq
\;\exp\left[-\frac{(q-\langle q \rangle_s)^2}{2A_{11}}\right]
\end{equation}

Furthermore with the transformation variables $h=\frac{q-\langle
q \rangle_s}{2A_{11}}$ , Eq.(\ref{66}) reduces to

\begin{equation}\label{67}
\chi(p_0, t) = \frac{1} {\pi^{1/2}} \int_{-\frac{\langle q(t)
\rangle_s}{[2 A_{11}(t)]^{1/2}}}^\infty dh\; e^{-h^2}
\end{equation}

The expression for $\kappa(t)$ is given by

\begin{equation}\label{68}
\kappa(t) = \frac{1} {\hbar \omega_0 (\bar{n}_0 + \frac{1} {2})}
\int_{-\infty}^{+\infty} dp_0 \;p_0\;
\exp\left[-\frac{\frac{p_0^2}{2}} {\hbar \omega_0 (\bar{n}_0 +
\frac{1} {2})} \right]\left[\frac{1} {\pi^{1/2}}
\int_{-\frac{\langle q(t) \rangle_s}{[2 A_{11}(t)]^{1/2}}}^\infty
dh \;e^{-h^2}\right]
\end{equation}

The primary quantities that determine $\kappa(t)$ are therefore
$\langle q(t) \rangle_s$ and $A_{11}(t)$ as given by Eq.(\ref{22})
and Eq.(\ref{28}), respectively, pertaining to the dynamics at the
barrier top. Putting the value of $\langle q(t) \rangle_s$ in the
lower limit of the second integral and carrying out integration by
parts we obtain the general expression for quantum transmission
coefficient $\kappa(t)$ as follows;

\begin{equation}\label{69}
\kappa (t) = \frac{C_v(t)} {\left\{ C_v^2(t) +
\frac{A_{11}(t)}{\hbar \omega_0 (\bar{n}_0 + \frac{1}
{2})}\right\}^{1/2}}\;\exp\left[-\;\frac{G^2(t)}{2\{C_v^2(t)\;
\hbar \omega_0 (\bar{n}_0 + \frac{1} {2}) + A_{11}(t) \}}\right]
\end{equation}

The expression Eq.(\ref{69}) is the direct quantum generalization
of Kohen-Tannor formulae \cite{tan} for classical time dependent
transmission coefficient. The classical formulae can be recovered
in the limit $k T \gg \hbar \omega_0$, so that $\hbar \omega_0
(\bar{n}_0 + \frac{1} {2})$ and $\hbar \omega_b (\bar{n}_b +
\frac{1} {2})$ reduce to $k T$ and $G(t)\rightarrow0$. A few
pertinent points are to be noted: First, since no approximation
has been made on the order of quantum correction in $G(t)$, the
expression takes care of quantum effects to all orders and thus
retains its validity even in the limit $T\rightarrow0$, i.e. , in
the vacuum regime. Second, since we have not assumed any specific
form of the density of modes of the bath oscillators so that
$\kappa (\omega) \rho (\omega)$ is arbitrary, the theory
developed so far is equipped to deal with arbitrary noise
correlation. Third, it is important to note that important
quantities in the expression Eq.(\ref{69}), $C_v(t)$,
$A_{11}(t)$, and $G(t)$ are the key elements required for
calculation of the dynamics of transmission coefficient
$\kappa(t)$. While $C_v(t)$ is the same as in the classical
theory, $A_{11}(t)$ and $G(t)$ are quantum mechanical objects.
$A_{11}(t)$ carries the imprints of quantum
fluctuation-dissipation relationship characterizing the quantum
nature of the heat bath. $G(t)$ on the other hand, as mentioned
earlier is a reflection of the characteristic of the nonlinear
potential. The interplay of the dissipation with the nonlinearity
of the system is expressed through this convolution integral. A
complete calculation of the transmission coefficient thus
requires the explicit knowledge of these three quantities.

In order to proceed further it is necessary to specify the nature
of the time dependent friction $\gamma(t)$. This is crucially
dependent on the distribution of the density of modes of the bath
oscillators. In the present case we assume a specific form of
Lorentian type, i.e. ;

\begin{equation}\label{70}
\kappa (\omega) \rho (\omega) = \frac{2}{\pi} \left(\frac{A}{1 +
\omega^2 \;\tau_c^2 }\right)
\end{equation}

where $A$ is the damping constant in the Markovian limit and
$\tau_c$ is the correlation time. This form of density by virtue
of Eq.(\ref{4}) results in,

\begin{eqnarray}
\gamma (t) & = & \frac{A}{\tau_c} e^{-t/\tau_c}\nonumber\\
& = & A\; \alpha\; e^{-\alpha t}\label{71}
\end{eqnarray}

where $\alpha=1/\tau_c$.

Depending on $\tau_c$ we consider two different limits; Markovian
and non-Markovian friction. We consider the former case first.

\subsection{Markovian friction}

For $\alpha\rightarrow\infty$ or $\tau_c\rightarrow0$, we have
$\kappa (\omega) \rho (\omega) = {2A}/\pi$; $\gamma(t)$ and its
Laplace transform thus reduce to

\begin{eqnarray*}
\gamma(t) = 2\;A \;\delta(t)
\end{eqnarray*}

and

\begin{equation}\label{72}
\tilde{\gamma}(\mu) = A
\end{equation}

respectively. According to Eq.(\ref{26}) $C_v(t)$ is then given by

\begin{equation}\label{73}
C_v(t) = \frac{e^{\mu_1 t } - e^{\mu_2 t }}{2 \omega_1}
\end{equation}

where $\mu_{1,2} = -\frac{A}{2} \pm \omega_1$ and $\omega_1 =
\sqrt{\frac{A^2}{4} + \omega_b^2}$. Furthermore the correlation
function as given by Eq.(\ref{11}) yields

\begin{equation}\label{74}
c(t_1 - t_2) = \frac{1}{2}\; A \;\hbar \omega_b \coth \left(
\frac{\hbar \omega_b}{2 k T}\right) \delta(t_1 - t_2)
\end{equation}

Making use of the last two relations for $C_v(t)$ and
$c(t_1-t_2)$ for the barrier top in Eq.(\ref{28}) we obtain

\begin{equation}\label{75}
A_{11}(t) = \frac{A}{2\omega_1^2}\; \hbar \omega_b (\bar{n}_b +
\frac{1} {2}) \left[\frac{1}{2\mu_1} e^{2\mu_1t} +
\frac{1}{2\mu_2} e^{2\mu_2t} +
\frac{2}{A}e^{-At}-\frac{2\omega_1^2}{\omega_b^2A}\right]
\end{equation}

Having known $C_v(t)$ and $A_{11}(t)$ in the Markovian limit we
are in a position to use them in the general expression for
quantum transmission coefficient Eq.(\ref{69}) to obtain,

\begin{eqnarray}\label{76}
\kappa (t) = \frac{e^{\mu_1 t } - e^{\mu_2 t }}{\left\{(e^{2\mu_1
t}+e^{2\mu_2 t}-2e^{-At}) + \frac{2 A \hbar \omega_b (\bar{n}_b +
\frac{1} {2})}{\hbar \omega_0 (\bar{n}_0 + \frac{1}
{2})}\left[\frac{1}{2\mu_1} e^{2\mu_1t} + \frac{1}{2\mu_2}
e^{2\mu_2t} +
\frac{2}{A}e^{-At}-\frac{2\omega_1^2}{\omega_b^2A}\right]\right\}^{1/2}}\nonumber\\
\times \exp\left[-\;\frac{G^2(t)}{\frac{\hbar \omega_0 (\bar{n}_0
+ \frac{1} {2})}{2\omega_1^2}\left[(e^{2\mu_1 t}+e^{2\mu_2
t}-2e^{-At}) + \frac{2 A \hbar \omega_b (\bar{n}_b + \frac{1}
{2})}{\hbar \omega_0 (\bar{n}_0 + \frac{1}
{2})}\left[\frac{1}{2\mu_1} e^{2\mu_1t} + \frac{1}{2\mu_2}
e^{2\mu_2t} +
\frac{2}{A}e^{-At}-\frac{2\omega_1^2}{\omega_b^2A}\right]\right]}\right]
\end{eqnarray}

It is instructive to work out the assymptotic limit of
$\kappa(t)$. For a bistable potential of the form
$V(q)=aq^4-bq^2$, the leading order quantum correction is given
by $Q(t)=-12\; a \;q(t)\; \langle\delta\hat{q}^2(t)\rangle$.
Since the dynamics pertains to the barrier top we have
approximately [see Appendix B]

\begin{equation}\label{77}
G(t) = -\frac{3 a}{\omega_b \;\omega_1\; A^2}\left[\hbar^3
\omega_0 (\bar{n}_0 + \frac{1}
{2})\right]^{1/2}\left[\frac{\omega_1}{\omega_b}\;\sinh{2\omega_bt}\;(e^{\mu_1
t } + e^{\mu_2 t }) - 2\cosh^2\omega_bt\;(e^{\mu_1 t } - e^{\mu_2
t })\right]
\end{equation}

which may be put in the form

\begin{equation}\label{78}
G(t) = -\frac{3 a}{\omega_b\; \omega_1\; A^2}\left[\hbar^3
\omega_0 (\bar{n}_0 + \frac{1} {2})\right]^{1/2}e^{\mu_1
t}\left[\frac{1}{2}\;e^{\ln\omega_1+2\omega_bt}-\frac{1}{2}\;e^{\ln\omega_b+2\omega_bt}-
\omega_b\right]
\end{equation}

in the long time for which $\ln\omega_1\ll2\omega_bt$ and
$\ln\omega_b\ll2\omega_bt$. Eq.(\ref{78}) then gives

\begin{equation}\label{79}
G(t) = \frac{3\; a}{\omega_b\; \omega_1\; A^2}\left[\hbar^3
\omega_0 (\bar{n}_0 + \frac{1} {2})\right]^{1/2}e^{\mu_1 t}
\end{equation}

Putting Eq.(\ref{79}) in Eq.(\ref{76}) and using the limit
$t\rightarrow\infty$ we obtain the assymptotic expression for the
transmission coefficient in the Markovian limit,

\begin{equation}\label{80}
\kappa(\infty) = \frac{1}{1 + \frac{A \;\hbar \omega_b (\bar{n}_b
+ \frac{1} {2})}{\mu_1\; \hbar \omega_0 (\bar{n}_0 + \frac{1}
{2})}}\exp\left[-\;\frac{18\; a^2\; \hbar^2\; \mu_1\;\omega_0
(\bar{n}_0 + \frac{1} {2})}{A^4\;
\omega_b^2\;\left(\mu_1\;\omega_0 (\bar{n}_0 + \frac{1} {2})+A
\;\omega_b (\bar{n}_b + \frac{1} {2})\right)}\right]
\end{equation}

In the classical limit as $\hbar\rightarrow\infty$, $\hbar
\omega_b (\bar{n}_b + \frac{1} {2})$ and $\hbar \omega_0
(\bar{n}_0 + \frac{1} {2})$ both goes over to $k T$; Eq.(\ref{80})
results in

\begin{equation}\label{81}
\kappa(\infty) = \left[\frac{1}{1+\frac{A}{\mu_1}}\right]^{1/2}
\end{equation}

Since $\mu_1+\mu_2=-A$ and $\mu_1\mu_2=- \omega_b^2$ the above
relation can be put into the following form

\begin{equation}\label{82}
\kappa(\infty) = \frac{\mu_1}{\omega_b}
\end{equation}

The classical Markovian coefficient, i.e. , the ratio of the
positive unstable root $\mu_1$ to the frequency of the barrier
$\omega_b$ is therefore reduced to Kramers' prefactor.

\subsection{Non-Markovian friction}

We now return to the Eq.(\ref{70}) and Eq.(\ref{71}).
Corresponding to Eq.(\ref{71}) we have
$\tilde{\gamma}(\mu)=\frac{A\;\alpha}{\mu+\alpha}$ from
Eq.(\ref{27}). $C_v(t)$ is therefore is given by

\begin{equation}\label{83}
C_v(t) = L^{-1}\left[\mu^2-\omega_b^2+\frac{\mu \;A \;\alpha}{\mu
+\alpha}\right]^{-1}
\end{equation}

Following Kohen and Tannor \cite{tan} we consider the case when
$\alpha$ is small, i.e. , the relevant zeros of the quantity in
the parenthesis in Eq.(\ref{83}) may be written as the solutions
of

\begin{equation}\label{84}
\mu^2-\omega_b^2+A \;\alpha = 0
\end{equation}

Depending on the relative magnitudes of $\omega_b^2$ and
$A\alpha$ it is convenient to discuss two situations
$\omega_b^2-A\alpha>0$ and $\omega_b^2-A\;\alpha<0$ separately.

(i) \underline{$\omega_b^2-A\;\alpha>0$}

Here the two roots become
$\mu_{1,2}=\pm\sqrt{\omega_b^2-A\;\alpha}$, with $\mu_1=-\mu_2$
and $C_v(t)$ is given by

\begin{equation}\label{85}
C_v(t) = \frac{\sinh\mu_1 t}{\mu_1}
\end{equation}

Since $\kappa
(\omega)\rho(\omega)=\frac{2A}{\pi}\frac{\alpha^2}{\omega^2+\alpha^2}$
we have from Eq.(\ref{11}) the correlation function as follows;

\begin{equation}\label{86}
c(t_1-t_2)=\frac{1}{2}\int_0^\infty
d\omega\left(\frac{2A}{\pi}\right)\frac{\alpha^2}{\omega^2+\alpha^2}\;\hbar
\omega\coth\left(\frac{\hbar
\omega}{2kT}\right)\cos\omega(t_1-t_2)
\end{equation}

Putting Eq.(\ref{85}) and Eq.(\ref{86}) in Eq. (\ref{28}) it is
possible to express $A_{11}(t)$ as

\begin{equation}\label{87}
A_{11}(t) = \frac{A\; \alpha^2\; \hbar \sinh^2 \mu_1 t}{ \pi
\mu_1^2} \;M - \frac{2 \;A\; \alpha^2\; \hbar \sinh \mu_1 t}{ \pi
\mu_1} \;N - \frac{2\; A\; \alpha^2\; \hbar \cosh \mu_1 t}{\pi}
\;O + \frac{A\; \alpha^2\; \hbar}{\pi}\;P
\end{equation}

where

\begin{eqnarray}
M & = & \int_0^\infty \frac{\omega\; \coth\frac{\hbar \omega}{2 k
T}}{(\alpha^2+\omega^2)\;(\omega^2+\mu_1^2)}\; d\omega\label{88}\\
N & = & \int_0^\infty \frac{\omega^2\; \sin\omega t
\;\coth\frac{\hbar \omega}{2 k
T}}{(\alpha^2+\omega^2)\;(\omega^2+\mu_1^2)^2}\; d\omega\label{89}\\
O & = & \int_0^\infty \frac{\omega\; \cos\omega t\;
\coth\frac{\hbar \omega}{2 k
T}}{(\alpha^2+\omega^2)\;(\omega^2+\mu_1^2)^2}\; d\omega\label{90}\\
P & = & \int_0^\infty \frac{\omega\; \coth\frac{\hbar \omega}{2 k
T}}{(\alpha^2+\omega^2)\;(\omega^2+\mu_1^2)^2}\; d\omega\label{91}
\end{eqnarray}

The nonlinear term $G(t)$ in Eq.(\ref{69}) can be obtained as
before by considering the leading order quantum correction as
shown in Appendix B for the case $\omega_b^2-A\alpha>0$. This is
given by

\begin{eqnarray}\label{92}
G(t)  =  -\frac{3\;a}{2\;\omega_b\;\mu_1^2 \;A\;\alpha}
\left[\hbar^3 \omega_0 \left(\bar{n}_0 + \frac{1} {2}\right)
\right]^{1/2} \left[\mu_1\; \sinh(\mu_1 t) +
\frac{A\;\alpha}{\omega_b}\cosh(\mu_1
t)\;\sinh(\frac{2\omega_b}{A}) \right. \nonumber\\
- \left. \omega_b\;\sinh(\frac{2 \omega_b}{A})\;\cosh(\mu_1
t-\frac{2 \mu_1}{A}) - \mu_1\;\cosh(\frac{2
\omega_b}{A})\;\sinh(\mu_1 t-\frac{2 \mu_1}{A})\right]
\end{eqnarray}

Having obtained $C_v(t)$, $A_{11}(t)$ and $G(t)$ from
Eq.(\ref{85}), Eq.(\ref{87}) and Eq.(\ref{92}), respectively, we
may write down the quantum transmission coefficient using
Eq.(\ref{69}) in the adiabatic regime ($\omega_b^2>A\alpha$)

\begin{eqnarray}\label{93}
\kappa(t) =\frac{[\frac{\sinh\mu_1
t}{\mu_1}]}{\left[\frac{\sinh^2\mu_1 t}{\mu_1^2} +
\frac{A\alpha^2\hbar}{\pi\hbar \omega_0 (\bar{n}_0 + \frac{1}
{2})}\left[\frac{\sinh^2\mu_1 t}{\mu_1^2}\;M -
\frac{2\sinh\mu_1 t}{\mu_1}\;N - 2\cosh\mu_1t\; O +2P \right]\right]^{1/2}}\nonumber\\
\times \exp\left[-\frac{G(t)^2}{2\left\{\hbar \omega_0 (\bar{n}_0
+ \frac{1} {2})\frac{\sinh^2\mu_1 t}{\mu_1^2} +
\frac{A\alpha^2\hbar}{\pi}(\frac{\sinh^2\mu_1 t}{\mu_1^2}\;M -
\frac{2\sinh\mu_1 t}{\mu_1}\;N - 2\cosh\mu_1t \;O
+2P)\right\}}\right]
\end{eqnarray}

In order to check the workability of the above method we first
test it against two typical benchmark calculations. To this end
we first employ the double-well potential as used earlier in
exact numerical calculation of quantum rate coefficient in
Ref.[22] with scaled parameters (actual parameters in Table I. of
DW1 in Ref.[22]) $E_b = 26.188$, $\omega_b = 1.0$, $\omega_0 =
1.414$ for two different scaled temperatures $kT = 2.617 (300K)$
and $kT= 1.744 (200K)$. The results for assymtotic transmission
coefficient as a function of damping constant from Fig.(9) of
Ref.[22] are compared against our analytical expression for the
same according to Eq.(\ref{93}) in Fig.(1). The agreement is
excellent at relatively higher temperature i.e. at $kT = 2.617
(300K)$. However it is apparent that at lower temperature the
analytical results is slightly higher than than exact one.
Keeping in view of the fact that the maximum deviation is less
than $8$ percent and that the analytical quantum correction in
Eq.(\ref{93}) is of lowest order, the result is fairly
satisfactory. For a more accurate calculation one would require
higher order corrections in $Q(t)$ which can be obtained from
numerical solution of the equations in Appendix A and be included
in $G(t)$ to be used in Eq.(\ref{63}). In what follows we also
make use of the classical results of Cohen and Tannor \cite{tan}
for a consistency check of our quantum calculation in the
approximate limits as we proceed. We now show the detailed
behavior of $\kappa(t)$ as a function of time for different
values of temperature $T$, damping constant $A$ and correlation
time $\tau_c(=1/\alpha)$. To this end we first compare the quantum
nature of transmission coefficient as a function of time against
its classical counterpart for $A=50$, $\tau_c=100$, $k T=1.0$,
$\omega_b=1.0$, $\omega_0=\sqrt{2}$ and $a=0.01$ in Fig.(2). It
is apparent that there is a significant increase in quantum
values. The variation of temperature has been depicted further in
Fig.(3). At higher temperature the quantum coefficient merges
into the classical coefficient of Kohen and Tannor \cite{tan}. An
important new content of the present theory is that it is
equipped to deal with zero point fluctuations. Around this
teperature the transmission coefficient hardly departs from its
maximum value. In Fig.(4) we show the explicit variation of
assymptotic transmission coefficient with temperature. While
classically $\kappa(t)$ is independent of temperature, quantum
nature of the heat bath imparts a temperature dependence which is
particularly strong in the low temperature regime. This is one of
the most distinctive features of the present calculation.

Fig.(5) shows the two sets of transmission coefficient $\kappa(t)$
as a function of time for different values of dissipation
parameters $A$ ranging from $30$ to $90$ one for quantum and
other for the classical. The input parameters are $k T=1.0$,
$\omega_b=1.0$, $\omega_0=\sqrt{2}$, $a=0.01$ and $\tau_c=100$.
It is apparent that there is marginal increase of the
corresponding quantum values when compared to the classical
transmission coefficient. In Fig.(6) the assymptotic transmission
coefficient is plotted as a function of damping constant. The
typical behavior in the intermediate to strong dissipation regime
is observed which is in complete conformity with the earlier
results. Fig.(7) shows an illustrative variation of the time
dependent quantum transmission coefficient for several values of
correlation time $\tau_c$. With increase in $\tau_c$ $\kappa(t)$
increases significantly - a trend which is similar to what is
observed in the classical case. With increase in time the
transmission coefficient $\kappa(t)$ reaches its plateau value in
the Markovian as well as non-adiabatic non-Markovian regime.

(ii) \underline{$\omega_b^2 - A\;\alpha < 0$}

Here from the equation for $\mu$, i.e. , $\mu^2 +
(A\;\alpha-\omega_b^2)=0$ we obtain two purely imaginary roots
$\mu_{2,3}$

\begin{eqnarray*}
\mu_{2,3} = \pm \;i \tilde{\mu}_2
\end{eqnarray*}

where $\tilde{\mu}_2=\sqrt{A\alpha-\omega_b^2}$. $C_v(t)$ in this
case is given by

\begin{equation}\label{94}
C_v(t) = \frac{\sin\tilde{\mu}_2 t}{\tilde{\mu}_2}
\end{equation}

Proceeding as before we arrive at the quantum transmission
coefficient in the so called caging regime as follows

\begin{eqnarray}\label{95}
\kappa(t) =\frac{[\frac{\sin\tilde{\mu}_2
t}{\mu_2}]}{\left[\frac{\sin^2\tilde{\mu_2} t}{\tilde{\mu_2}^2} +
\frac{A\alpha^2\hbar}{\pi\hbar \omega_0 (\bar{n}_0 + \frac{1}
{2})}\left[\frac{\sin^2\tilde{\mu}_2 t}{\tilde{\mu}_2^2}\;M_1 -
\frac{2\sin\mu_2 t}{\tilde{\mu}_2}\;N_1 - 2\cos\tilde{\mu}_2t\;
O_1 +2P_1 \right]\right]^{1/2}}
\nonumber\\
\times \exp\left[-\frac{G(t)^2}{2\left\{\hbar \omega_0 (\bar{n}_0
+ \frac{1} {2})\frac{\sin^2\mu_2 t}{\tilde{\mu}_2^2} +
\frac{A\alpha^2\hbar}{\pi}(\frac{\sin^2\tilde{\mu}_2
t}{\tilde{\mu}_2^2}\;M_1 - \frac{2\sin\tilde{\mu}_2
t}{\tilde{\mu}_2}\;N_1 - 2\cos\tilde{\mu}_2t \;O_1
+2P_1)\right\}}\right]
\end{eqnarray}

Here $G(t)$ is given by [see Appendix B]

\begin{eqnarray}\label{96}
G(t)  =  -\frac{3\;a}{2\;\omega_b\;\tilde{\mu}_2^2 \;A\;\alpha}
\left[\hbar^3 \omega_0 \left(\bar{n}_0 +
\frac{1}{2}\right)\right]^{1/2} \left[ \omega_b\;
\sinh\left(\frac{2 \omega_b}{A}\right)\;\cos\left(\tilde{\mu}_2
t-\frac{2\;\tilde{\mu}_2}{A} \right)
\right. \nonumber\\
-\left. \tilde{\mu}_2\;\cosh\left(\frac{2 \omega_b}{A}\right)\;
\sin\left(\tilde{\mu}_2 t -\frac{2 \tilde{\mu}_2}{A}\right) +
\tilde{\mu}_2\;\sin\left( \tilde{\mu}_2 t\right) -
\frac{A\alpha}{\omega_b}\cos\left(\tilde{\mu}_2
t\right)\;\sinh\left(\frac{2\omega_b}{A}\right)\right]
\end{eqnarray}

where

\begin{eqnarray}
M_1 & = & \int_0^\infty \frac{\omega\; \coth\frac{\hbar \omega}{2
k
T}}{(\alpha^2+\omega^2)\;(\omega^2-\tilde{\mu}_2^2)}\; d\omega\label{97}\\
N_1 & = & \int_0^\infty \frac{\omega^2\; \sin\omega t
\;\coth\frac{\hbar \omega}{2 k
T}}{(\alpha^2+\omega^2)\;(\omega^2-\tilde{\mu}_2^2)^2}\; d\omega\label{98}\\
O_1 & = & \int_0^\infty \frac{\omega\; \cos\omega t\;
\coth\frac{\hbar \omega}{2 k
T}}{(\alpha^2+\omega^2)\;(\omega^2-\tilde{\mu}_2^2)^2}\; d\omega\label{99}\\
P_1 & = & \int_0^\infty \frac{\omega\; \coth\frac{\hbar \omega}{2
k T}}{(\alpha^2+\omega^2)\;(\omega^2-\tilde{\mu}_2^2)^2}\;
d\omega\label{100}
\end{eqnarray}

The behavior of the transmission coefficient is very much similar
to that for the classical case as discussed in detail by Kohen
and Tannor \cite{tan}. In order to allow ourselves a fair
comparison with this theory we calculate according to
Eq.(\ref{95}) and the corresponding classical case for the
parameters $A=150$, $\tau_c=100$, $k T=1.0$, $\omega_b=1.0$,
$\omega_0=\sqrt{2}$ and $a=0.01$. The results are shown in
Fig.(8). It is apparent that the oscillating nature persists even
in the quantum case although this is accompanied by a loss of
amplitude compared to the classical case. In Fig.$(9)$ and
Fig.$(10)$ we illustrate the dependence of transmission
coefficient with temperature and damping, respectively. Apart
from temperature dependence no significant difference from
classical behaviour is observed.

\section{Conclusion}

In this paper we have derived a general time-dependent quantum
Kramers'-Grote-Hynes rate constant for a chemical reaction in a
condensed medium. This is based on quantum phase space probability
distribution functions proposed recently by us to describe
quantum Brownian motion in terms of a generalized quantum
Langevin equation. We have correlated the time dependence of the
transmission coefficient in detail with the variation of
dissipation, temperature and correlation time of the noise. Our
results can be summarized as follows:

(i) The quantum KGH transmission coefficient Eq.(\ref{69}), the
central result of this paper reduces to to its well known
classical limit when $\hbar\rightarrow0$. The expression retains
its full validity even in the vacuum limit, i.e. ,
$T\rightarrow0$.

(ii) We have worked out in detail the specific form of this
transmission coefficient in the Markovian as well as in the
non-Markovian case. Depending on the dissipation, correlation
time and frequency at the barrier top, both adiabatic and
non-adiabatic situations have been considered with a comparison
with the classical theory of Kohen and Tannor \cite{tan}.While
classical transmission coefficient is independent of temperature
its quantum counterpart depends on temperature significantly,
particularly in the low temperature regime marked by quantum
effects.

(iii) The general expression for quantum KGH transmission
coefficient contains quantum corrections to all orders. We have
shown that the systematic corrections can be obtained explicitly
order by order by simply solving a number of coupled ordinary
differential equations.

(iv) The expression for the quantum KGH transmission coefficient
is a natural extension of classical theory to quantum domain and
provides a unified description of thermally activated process and
tunneling.

Before conclusion we now make a few remarks on the present method.

First, since the advent of quantum mechanics several methods of
quantization have been in use for description of quantum Brownian
motion. Of these, Lagrangian and Hamiltonian formulation have
been proposed in the framework of Feynman (path integral)
quantization \cite{weiss,calde,leg} and been most widely used
since early eighties. Dynamical quantization \cite{bol,boli,lsf}
have been tried with some preliminary success. The present method
is based on canonical Heisenberg quantization procedure and has
been recently utilized to propose non-Markovian quantum version of
Kramers \cite{db1,db2}, Fokker-Planck \cite{skb}, Smoluchowski
\cite{db3} and diffusion equations \cite{skb}. The main advantage
of this approach is that it allows the classical method of
solution of equation for probability distribution functions to be
applied in the same way as one makes use of flux-over-population
method or reactive flux method for calculation of rate
coefficients.

Second, since the phase space function formulation of classical
reactive flux theory by Kohen and Tannor is particularly useful
for visualization of the behaviour of phase space trajectories, a
direct extension of this method to the quantum domain as done in
the present formulation allows us to achieve a similar objective
in terms of c-number trajectories. It is also important to point
out that path integral approach also gives rise to several
approximations that have a classical flavour \cite{voth,wsmi}.

Third, an important advantage of this approach is that one not
only obtains the asymptotic quantum rate coefficient but also the
details of its behaviour at all time \textit{analytically}
including its approach towards the steady state. In this sense it
is complimentary to the exact numerical work of Ref.[22].

Fourth, although the present method employs system-reservoir
oscillator model for evolution of quantum mechanical rate, in the
phase space distribution function approach bath co-ordinates do
not appear explicitly in the ultimate calculation. One can also
avoid multidimensional integration employed in more powerful path
integral approach. The present method instead takes care of
improvement of quantum corrections successively order by order by
solving coupled ordinary differential equations of Appendix (A).

In the view of the above discussions we anticipate that the
present approach can readily be adopted to numerical scheme, for
example, to that of classical method of Berne and co-workers
\cite{jes,jest}, for calculation of time- dependent quantum
transmission coefficient beyond linearization over the entire
range of friction, noise correlation and temperature, to explain
quantum turnover and other interesting features. The work in this
direction is in progress.

\begin{acknowledgments}
The authors are indebted to the Council of Scientific and
Industrial Research for partial financial support under Grant No.
01/(1740)/02/EMR-II. SKB expresses his sincerest gratitude to
MPI-PKS for kind hospitality through Visiting Scientist Program.
\end{acknowledgments}

\begin{appendix}

\section{Evolution Equations For Higher-Order Quantum Corrections For
Anharmonic Potential}

\noindent The equations upto fourth order for quantum corrections
(corresponding to the contribution of anharmonicity of the
potential) are listed below.

\noindent Equations for the second order are:

\begin{eqnarray}
\frac{d}{dt} \langle \delta \hat{q}^2 \rangle &=& \langle \delta
\hat{q} \delta \hat{p} +
\delta \hat{p} \delta \hat{q} \rangle,                 \nonumber  \\
\frac{d}{dt} \langle \delta \hat{p}^2 \rangle &=& -2\Gamma \langle
\delta \hat{p}^2 \rangle -V^{\prime\prime} \langle \delta \hat{q}
\delta \hat{p} + \delta \hat{p} \delta \hat{q} \rangle -
V^{\prime\prime\prime} \langle \delta \hat{q} \delta \hat{p}
\delta \hat{q} \rangle,
\label{eqnB1}  \\
\frac{d}{dt} \langle \delta \hat{q} \delta \hat{p} + \delta
\hat{p} \delta \hat{q} \rangle &=& -\Gamma \langle \delta \hat{q}
\delta \hat{p} + \delta \hat{p} \delta \hat{q} \rangle 2\langle
\delta \hat{p}^2 \rangle - 2V^{\prime\prime} \langle \delta
\hat{q}^2 \rangle - V^{\prime\prime\prime} \langle \delta
\hat{q}^3 \rangle, \nonumber
\end{eqnarray}

\noindent Those for the third order are:

\begin{eqnarray}
\frac{d}{dt} \langle \delta \hat{q}^3 \rangle &=&
3\langle \delta \hat{q} \delta \hat{p} \delta \hat{q} \rangle,  \nonumber \\
\frac{d}{dt} \langle \delta \hat{p}^3 \rangle &=& -3\Gamma \langle
\delta \hat{p}^3 \rangle -3V^{\prime\prime} \langle \delta \hat{p}
\delta \hat{q} \delta \hat{p} \rangle + V^{\prime\prime\prime}
\left( \frac{3}{2} \langle \delta \hat{q}^2 \rangle \langle \delta
\hat{p}^2 \rangle - \frac{3}{2}
\langle \delta \hat{p} \delta \hat{q}^2 \delta \hat{p} \rangle + \hbar^2 \right), \nonumber \\
\frac{d}{dt} \langle \delta \hat{q} \delta \hat{p} \delta \hat{q}
\rangle &=& -\Gamma \langle \delta \hat{q} \delta \hat{p} \delta
\hat{q} \rangle + 2\langle \delta \hat{p} \delta \hat{q} \delta
\hat{p} \rangle - V^{\prime\prime} \langle \delta \hat{q}^3
\rangle - \frac{V^{\prime\prime\prime}}{2} \left( \langle \delta
\hat{q}^4 \rangle -
{\langle \delta \hat{q}^2 \rangle}^2 \right),                 \label{eqnB2}  \\
\frac{d}{dt} \langle \delta \hat{p} \delta \hat{q} \delta \hat{p}
\rangle &=& -2\Gamma \langle \delta \hat{p} \delta \hat{q} \delta
\hat{p} \rangle + \langle \delta \hat{p}^3 \rangle -
2V^{\prime\prime}
\langle \delta \hat{q} \delta \hat{p} \delta \hat{q} \rangle  \nonumber \\
&+& \frac{V^{\prime\prime\prime}}{2} \left( \langle \delta
\hat{q}^2 \rangle \langle \delta \hat{q} \delta \hat{p} + \delta
\hat{p} \delta \hat{q} \rangle - \langle \delta \hat{q}^3 \delta
\hat{p} + \delta \hat{p} \delta \hat{q}^3 \rangle \right),
\nonumber
\end{eqnarray}

\noindent And lastly, the fourth order equations are:

\begin{eqnarray}
\frac{d}{dt} \langle \delta \hat{q}^4 \rangle &=&
2\langle \delta \hat{q}^3 \delta \hat{p} + \delta \hat{p} \delta \hat{q}^3 \rangle,  \nonumber \\
\frac{d}{dt} \langle \delta \hat{p}^4 \rangle &=& -4\Gamma \langle
\delta \hat{p}^4 \rangle -2V^{\prime\prime} \langle \delta \hat{q}
\delta \hat{p}^3 + \delta \hat{p}^3 \delta \hat{q} \rangle +
2V^{\prime\prime\prime} \langle \delta \hat{q}^2 \rangle \langle
\delta \hat{p}^3 \rangle,
\nonumber  \\
\frac{d}{dt} \langle \delta \hat{q}^3 \delta \hat{p} + \delta
\hat{p} \delta \hat{q}^3 \rangle &=& -\Gamma \langle \delta
\hat{q}^3 \delta \hat{p} + \delta \hat{p} \delta \hat{q}^3 \rangle
-2V^{\prime\prime} \langle \delta \hat{q}^4 \rangle - 3\hbar^2 +
6\langle \delta \hat{p} \delta \hat{q}^2 \delta \hat{p} \rangle   \nonumber   \\
&+& V^{\prime\prime\prime} \langle \delta \hat{q}^2 \rangle
\langle \delta \hat{q}^3 \rangle,
\label{eqnB3}   \\
\frac{d}{dt} \langle \delta \hat{q} \delta \hat{p}^3 + \delta
\hat{p}^3 \delta \hat{q} \rangle &=& -3\Gamma \langle \delta
\hat{q} \delta \hat{p}^3 + \delta \hat{p}^3 \delta \hat{q} \rangle
+ 2\langle \delta \hat{p}^4 \rangle + 3V^{\prime\prime} (\hbar^2 -
2\langle \delta \hat{p} \delta \hat{q}^2 \delta \hat{p} \rangle)   \nonumber  \\
&+& 3V^{\prime\prime\prime} \langle \delta \hat{q}^2 \rangle
\langle \delta \hat{p} \delta \hat{q} \delta \hat{p} \rangle,
\nonumber  \\
\frac{d}{dt} \langle \delta \hat{p} \delta \hat{q}^2 \delta
\hat{p} \rangle &=& -2\Gamma \langle \delta \hat{p} \delta
\hat{q}^2 \delta \hat{p} \rangle -V^{\prime\prime} \langle \delta
\hat{q}^3 \delta \hat{p} + \delta \hat{p} \delta \hat{q}^3 \rangle
+ \langle \delta \hat{p}^3 \delta \hat{q} + \delta \hat{q} \delta \hat{p}^3 \rangle    \nonumber \\
&+& V^{\prime\prime\prime} \langle \delta \hat{q}^2 \rangle
\langle \delta \hat{q} \delta \hat{p} \delta \hat{q} \rangle.
\nonumber
\end{eqnarray}

\noindent The derivatives of $V(q)$, i.e., $V^{\prime\prime}$ or
$V^{\prime\prime\prime}$ etc. in the above expressions are
functions of $q$ the dynamics of which is given by Eq.(\ref{12}).

\section{Calculation of $G(t)$}

It is already been pointed out that $G(t)$ as defined in
Eq.(\ref{24}) describes an interplay of dissipation and
nonlinearity in terms of the following convolution integral

\begin{equation}\label{A1}
G (t) = \int_0^t C_v(t-\tau)\; Q(\tau) d\tau
\end{equation}

where $Q(t)$ is given by Eq.(\ref{15}) and the relaxation
function $C_v(t)$ refers to Eq.(\ref{26}) pertaining to the
dynamics at the barrier top. When $Q(t)$ is taken in full $G(t)$
contains quantum corrections to all orders and is formally exact.
However for practical calculation we evaluate this quantity order
by order. Leading order quantum correction in Eq.(\ref{15}) as
given by $Q(t)=-\frac{1} {2} V'''(q) \langle \delta \hat{q} (t)^2
\rangle$ for the bistable potential $V(q)=a\;q^4-b\;q^2$ results
in

\begin{equation}\label{A2}
Q(t) = -12\;a\;q(t)\;\langle \delta \hat{q} (t)^2 \rangle
\end{equation}

Explicit solution of $\langle \delta \hat{q} (t)^2 \rangle$ at
the barrier top \cite{db1,sm} at $q=q_0=0$ is given by$\langle
\delta \hat{q} (t)^2 \rangle=\langle \delta \hat{q}^2
\rangle_{t=0}\;\cosh2\omega_bt + \frac{\langle\delta \hat{q}\;
\delta \hat{p} + \delta \hat{p}\; \delta
\hat{q}\rangle_{t=0}}{2\omega_b}\;\sinh2 \omega_b t$. Furthermore
$q(t)$ is known from Eq.(\ref{20}) upto a leading order so that
we have

\begin{equation}\label{A3}
q(t) = p_0\; C_v(t)
\end{equation}

Since with minimum uncertainty product state \cite{db1,sm} we
require

\begin{equation}\label{A4}
\langle\delta \hat{q}\; \delta \hat{p} + \delta \hat{p}\; \delta
\hat{q}\rangle_{t=0}\;=\;0
\end{equation}

\begin{equation}\label{A5}
\langle \delta \hat{q}^2 \rangle_{t=0} = \frac{\hbar}{2 \omega_b}
\end{equation}

and so we have

\begin{equation}\label{A6}
\langle \delta \hat{q} (t)^2 \rangle = \frac{\hbar}{2
\omega_b}\;\cosh 2\omega_b t
\end{equation}

For an initially thermally distributed $p_0$ we may write

\begin{equation}\label{A7}
p_0 = \left[\hbar \omega_0 (\bar{n}_0 + \frac{1} {2})\right]^{1/2}
\end{equation}

With Eq.(\ref{A7}), Eq.(\ref{A6}) and Eq.(\ref{A3}), $Q(t)$ yields

\begin{equation}\label{A8}
Q(t) = -\frac{6\;a\;\hbar^3 \omega_0 (\bar{n}_0 + \frac{1}
{2})}{\omega_b}\;C_v(t)\;\cosh 2\omega_b t
\end{equation}

Putting Eq.(\ref{A8}) in Eq.(\ref{A1}) we obtain the general
expression for $G(t)$ upto the leading order in quantum
correction as

\begin{equation}\label{A9}
G(t) = -\frac{6\;a}{\omega_b}\left[\hbar^3 \omega_0 (\bar{n}_0 +
\frac{1} {2})\right]^{1/2}\int_0^tC_v(t-\tau)\;C_v(\tau)
\cosh2\omega_b\tau \;d\tau
\end{equation}

The rest of the treatment depends on specific nature of $C_v(t)$.

Case(I) Markovian friction: For this case $C_v(t)$ is given by
Eq.(\ref{73}). Explicit evaluation of the integral in
Eq.(\ref{A9}) results

\begin{equation}\label{A10}
G(t) = -\frac{3\;a}{\omega_1\; \omega_b\; A^2 }\left[\hbar^3
\omega_0 (\bar{n}_0 + \frac{1}
{2})\right]^{1/2}\left[\frac{\omega_1}{\omega_b}\;\sinh 2\omega_b
t\;(e^{\mu_1 t}+e^{\mu_2 t})-2\;\cosh^2\omega_b t \;(e^{\mu_1
t}-(e^{\mu_2 t})\right]
\end{equation}

Case(II) non-Markovian friction:

Similarly the evaluation of the integral in Eq.(\ref{A9}) using
$C_v(t)$ from Eq.(\ref{85}) for the non-adiabatic case
$\omega_b^2-A\alpha>0$ leads to

\begin{eqnarray}\label{A11}
G(t)  =  -\frac{3\;a}{2\;\omega_b\;\mu_1^2 \;A\;\alpha}
\left[\hbar^3 \omega_0 \left(\bar{n}_0 + \frac{1} {2}\right)
\right]^{1/2} \left[\mu_1\; \sinh(\mu_1 t) +
\frac{A\;\alpha}{\omega_b}\cosh(\mu_1
t)\;\sinh(\frac{2\omega_b}{A}) \right. \nonumber\\
- \left. \omega_b\;\sinh(\frac{2 \omega_b}{A})\;\cosh(\mu_1
t-\frac{2 \mu_1}{A}) - \mu_1\;\cosh(\frac{2
\omega_b}{A})\;\sinh(\mu_1 t-\frac{2 \mu_1}{A})\right]
\end{eqnarray}

where to avoid divergence we have used a cut off for the integral
at time $t=1/A$. This is a good approximation in view of the fact
that quantum correction in Eq.(\ref{A11}) has been considered to
a leading order.

For adiabatic or so called caging regime ($\omega_b^2 -
A\alpha<0$), on the other hand $C_v(t)$ is given by Eq.(\ref{94}).
The corresponding $G(t)$ can be calculated to obtain from
Eq.(\ref{A9}) as

\begin{eqnarray}\label{A12}
G(t)  =  -\frac{3\;a}{2\;\omega_b\;\tilde{\mu}_2^2 \;A\;\alpha}
\left[\hbar^3 \omega_0 \left(\bar{n}_0 +
\frac{1}{2}\right)\right]^{1/2} \left[ \omega_b\;
\sinh\left(\frac{2 \omega_b}{A}\right)\;\cosh\left(\tilde{\mu}_2
t-\frac{2\;\tilde{\mu}_2}{A} \right)
\right. \nonumber\\
-\left. \tilde{\mu}_2\;\cosh\left(\frac{2 \omega_b}{A}\right)\;
\sinh\left(\tilde{\mu}_2 t-\frac{2 \tilde{\mu}_2}{A}\right) +
\tilde{\mu}_2\;\sinh\left( \tilde{\mu}_2 t\right) -
\frac{A\alpha}{\omega_b}\cosh\left(\tilde{\mu}_2
t\right)\;\sinh\left(\frac{2\omega_b}{A}\right)\right]
\end{eqnarray}

\end{appendix}

\newpage

\begin{figure}[h]
\includegraphics[width=0.7\linewidth,angle=-90]{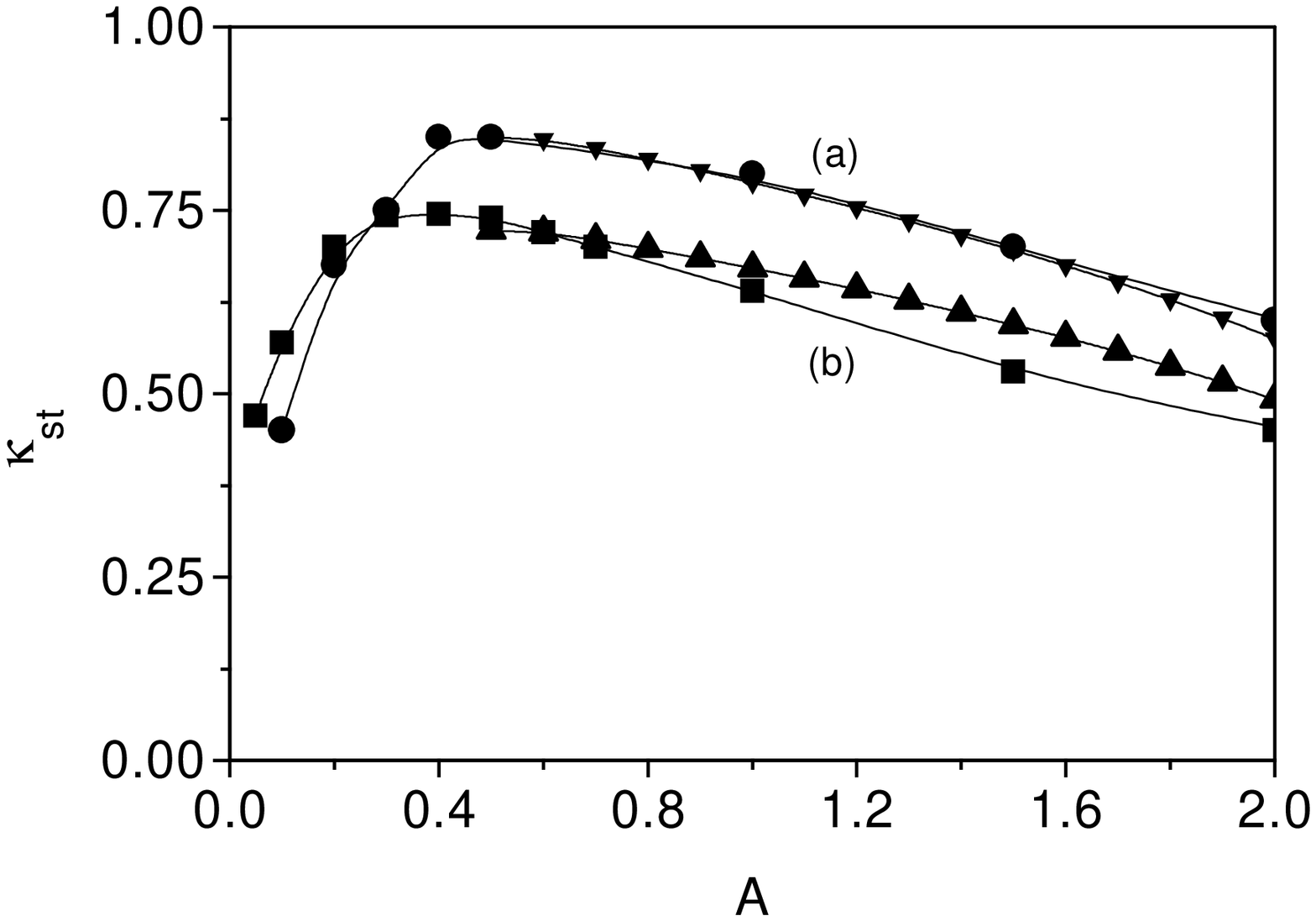}
\vspace*{1.0cm}
\caption{
Asymptotic quantum transmission coefficient $\kappa_{st}$
is plotted as a function of damping strength (A) at two different
temperatures (a) $kT=2.617 (300K)$ and (b) $kT=1.744(200K)$ for
the parameter set mentioned in the text. The numerical results of
Fig.9 of Ref.(22) are shown by solid circles and squares and
compared to our analytical results (solid triangles)
corresponding to Eq.(93).
}
\end{figure}


\begin{figure}[h]
\includegraphics[width=0.7\linewidth,angle=-90]{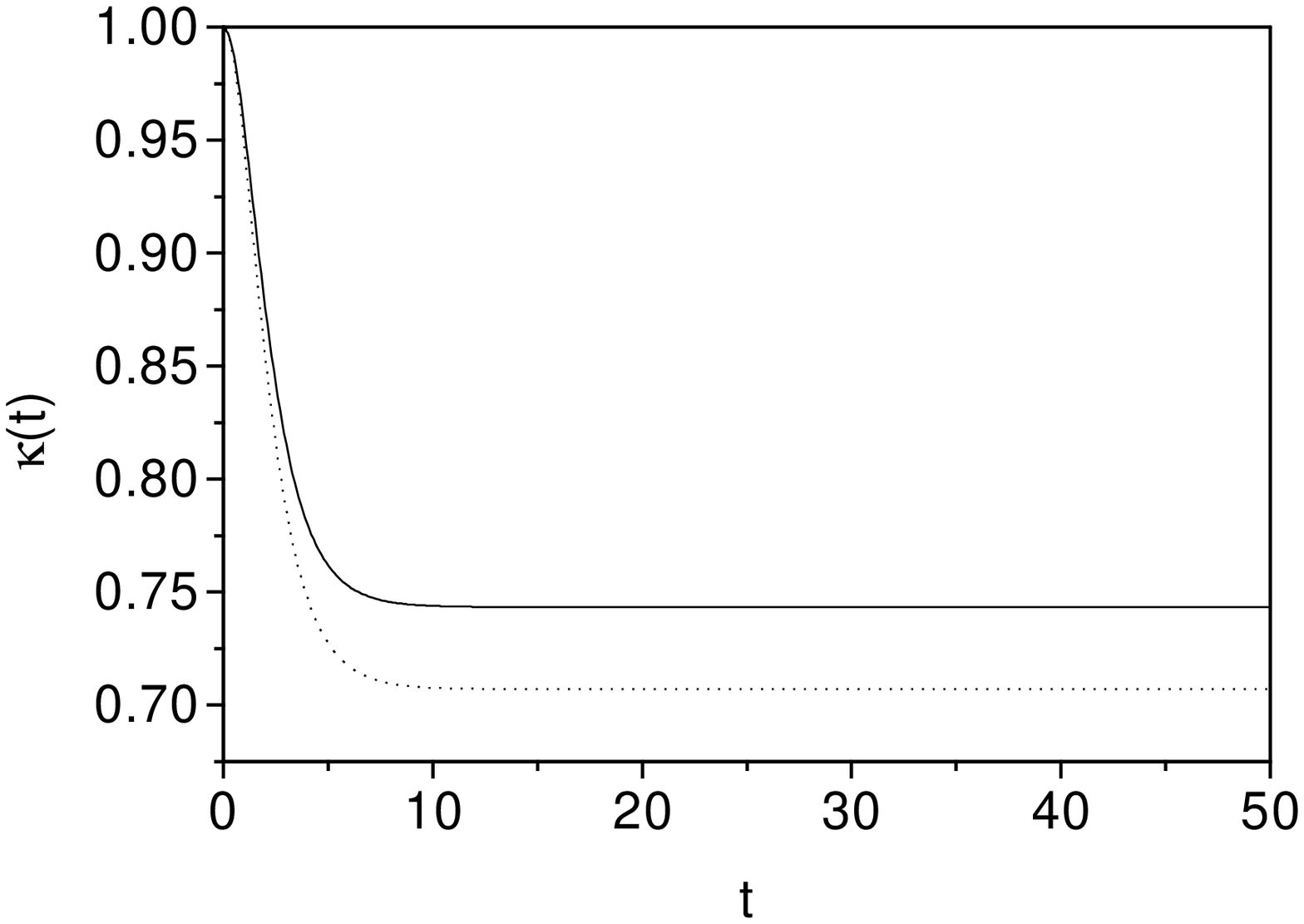}
\vspace*{0.5cm}
\caption{
The quantum (continuous line) transmission coefficient is plotted against
time and compared to classical (dotted line) one for the parameter values
$A=50$, $\tau_c=100$, $kT=1.0$, $\omega_b=1.0$, $\omega_0=\sqrt{2}$, $a=0.01$.
}
\end{figure}

\newpage

\begin{figure}[h]
\includegraphics[width=0.7\linewidth,angle=-90]{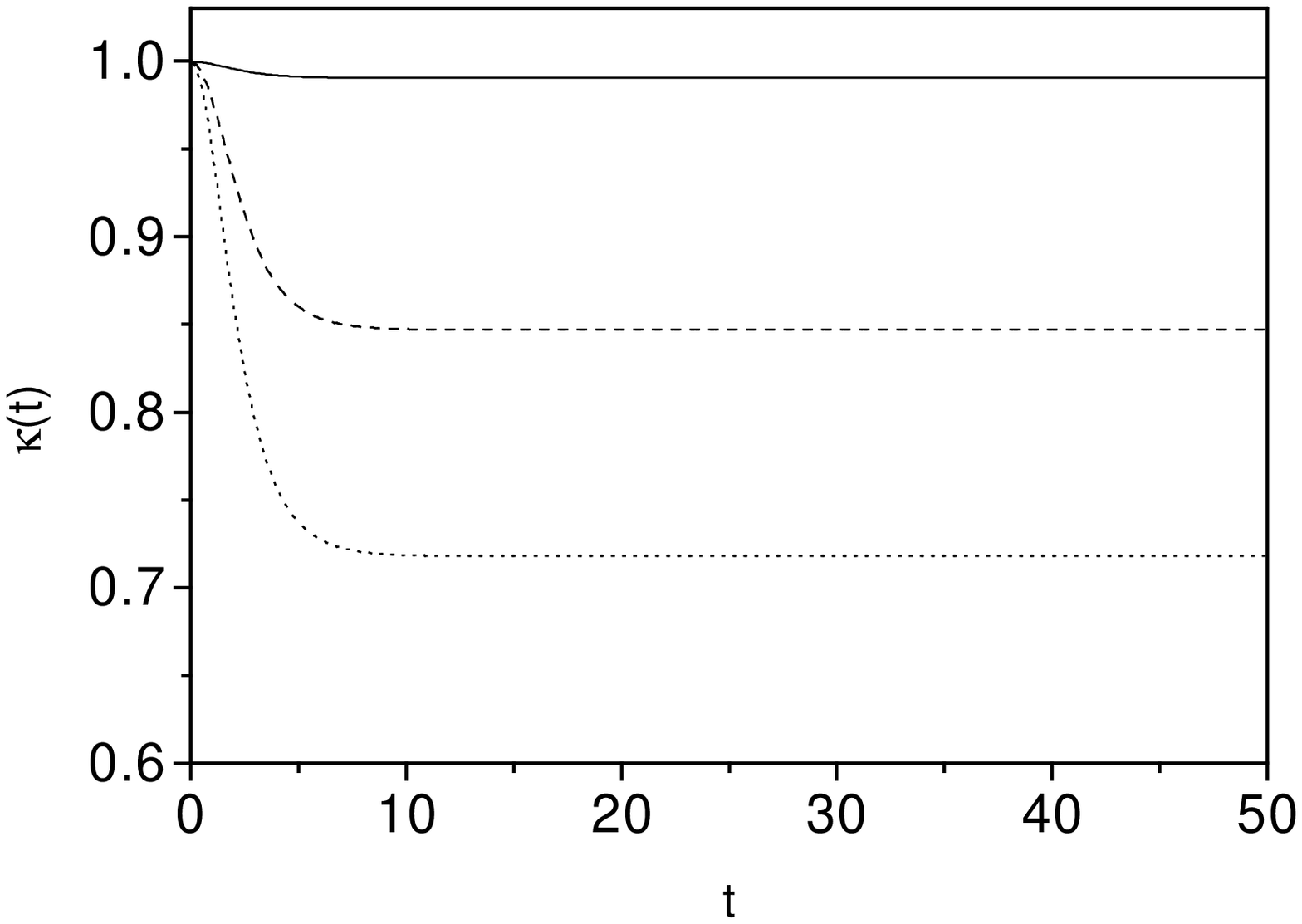}
\vspace*{0.5cm}
\caption{
The quantum transmission coefficient is plotted against time for several
values of temperature $kT=0.0$ (continuous line); $kT=0.3$ (dashed line);
$kT=25.0$ (dotted line) for $A=50$, $\tau_c=100$, $kT=1.0$, $\omega_b=1.0$,
$\omega_0=\sqrt{2}$, $a=0.01$.
}
\end{figure}


\begin{figure}[h]
\includegraphics[width=0.7\linewidth,angle=-90]{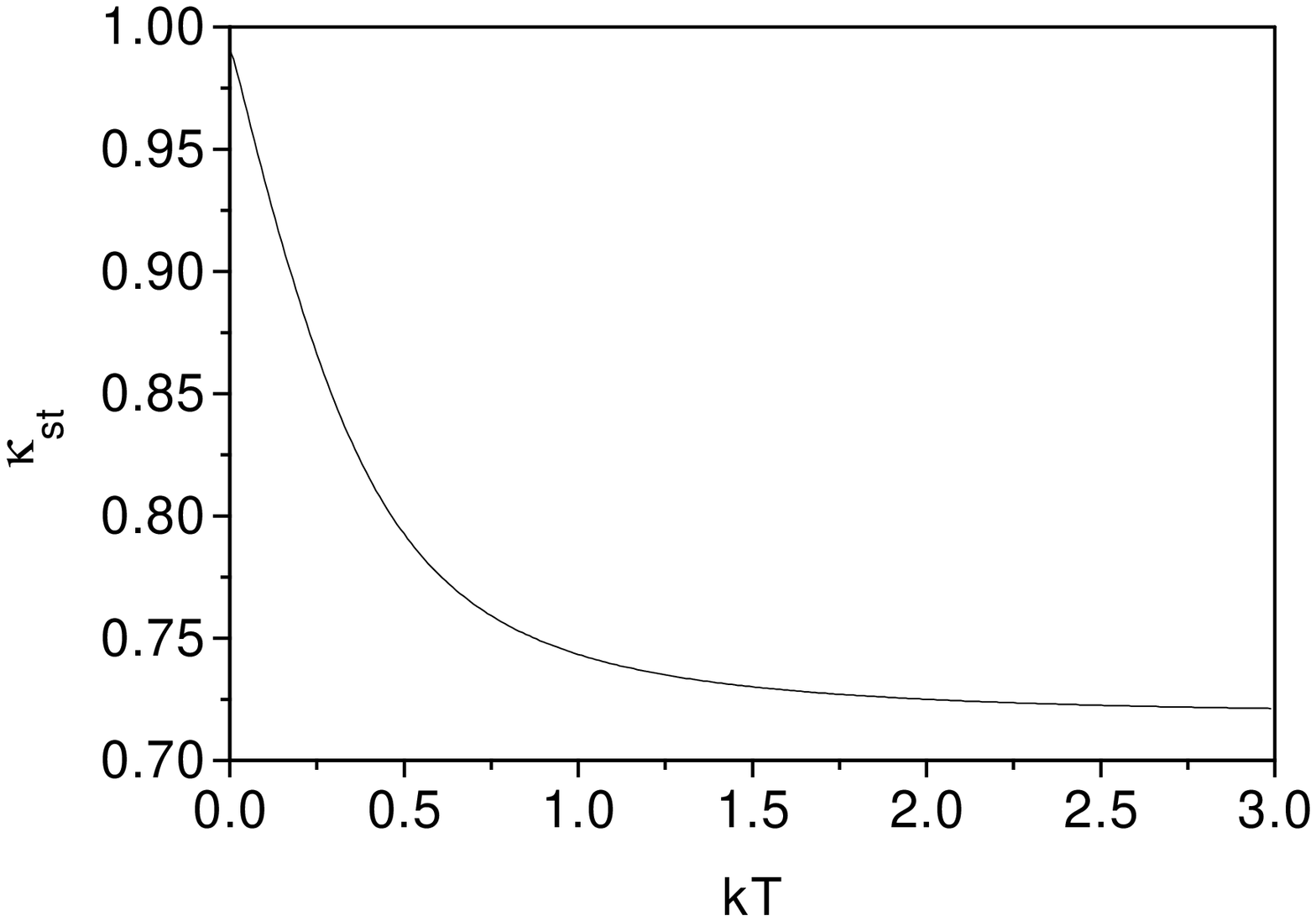}
\vspace*{0.5cm}
\caption{
The assymptotic quantum transmission coefficient $\kappa_{st}$ is
plotted against temperature $kT$ for the set of parameters as in Fig.2.
}
\end{figure}

\newpage

\begin{figure}[h]
\includegraphics[width=0.7\linewidth,angle=-90]{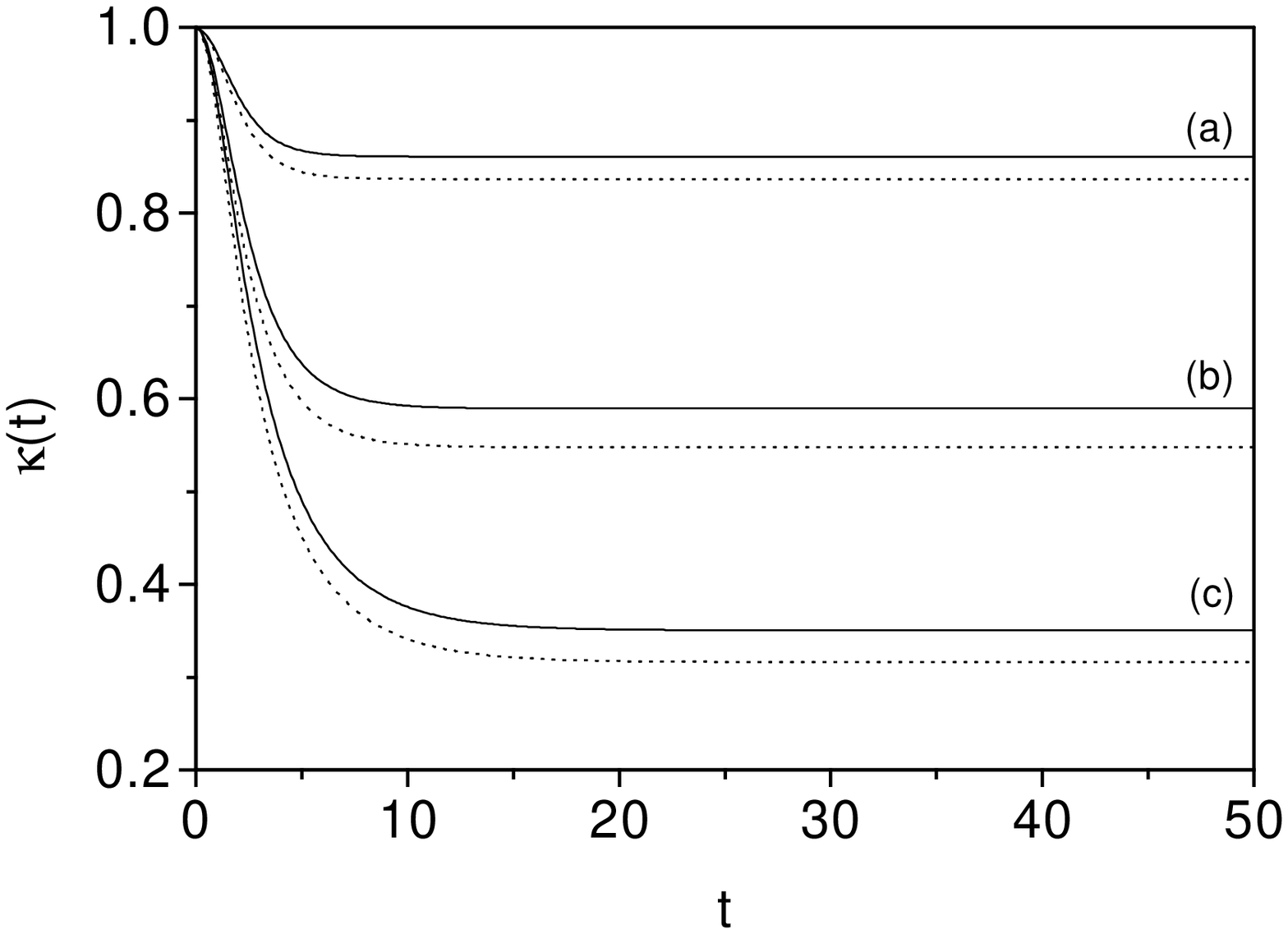}
\vspace*{0.5cm}
\caption{
The quantum (continuous lines) transmission coefficient $\kappa(t)$
is plotted against time for different values of damping constant (a) $A=30$,
(b) $A=70$, (c) $A=90$ and compared with classical (dotted lines) ones for
the parameters $\tau_c=100$, $kT=1.0$, $\omega_b=1.0$, $\omega_0=\sqrt{2}$,
$a=0.01$.
}
\end{figure}


\begin{figure}[h]
\includegraphics[width=0.7\linewidth,angle=-90]{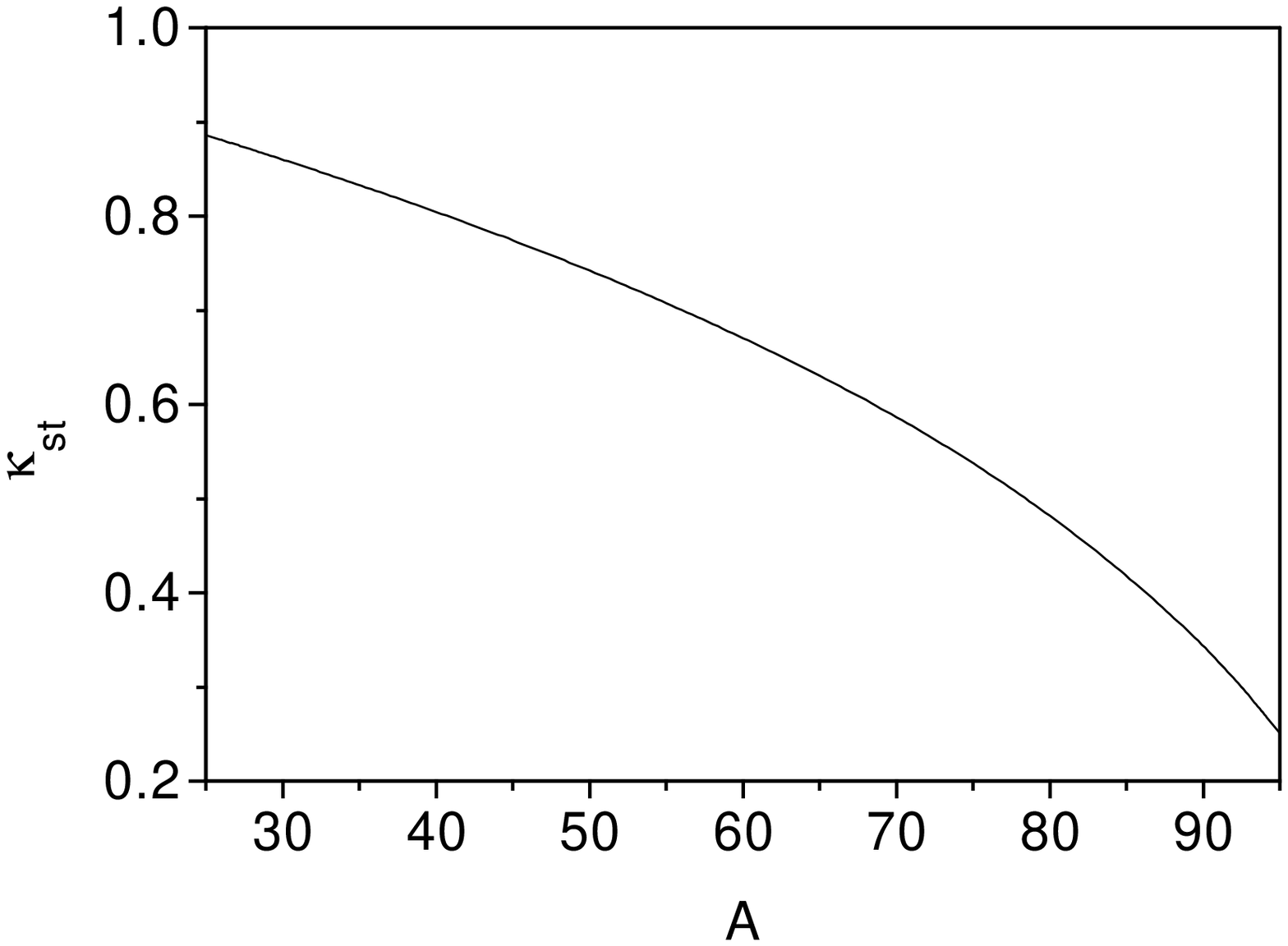}
\vspace*{0.5cm}
\caption{
The assymptotic quantum transmission coefficient $\kappa(t)$
is plotted against
damping constant $A$ for the parameter values as in Fig.4.
}
\end{figure}

\newpage

\begin{figure}[h]
\includegraphics[width=0.7\linewidth,angle=-90]{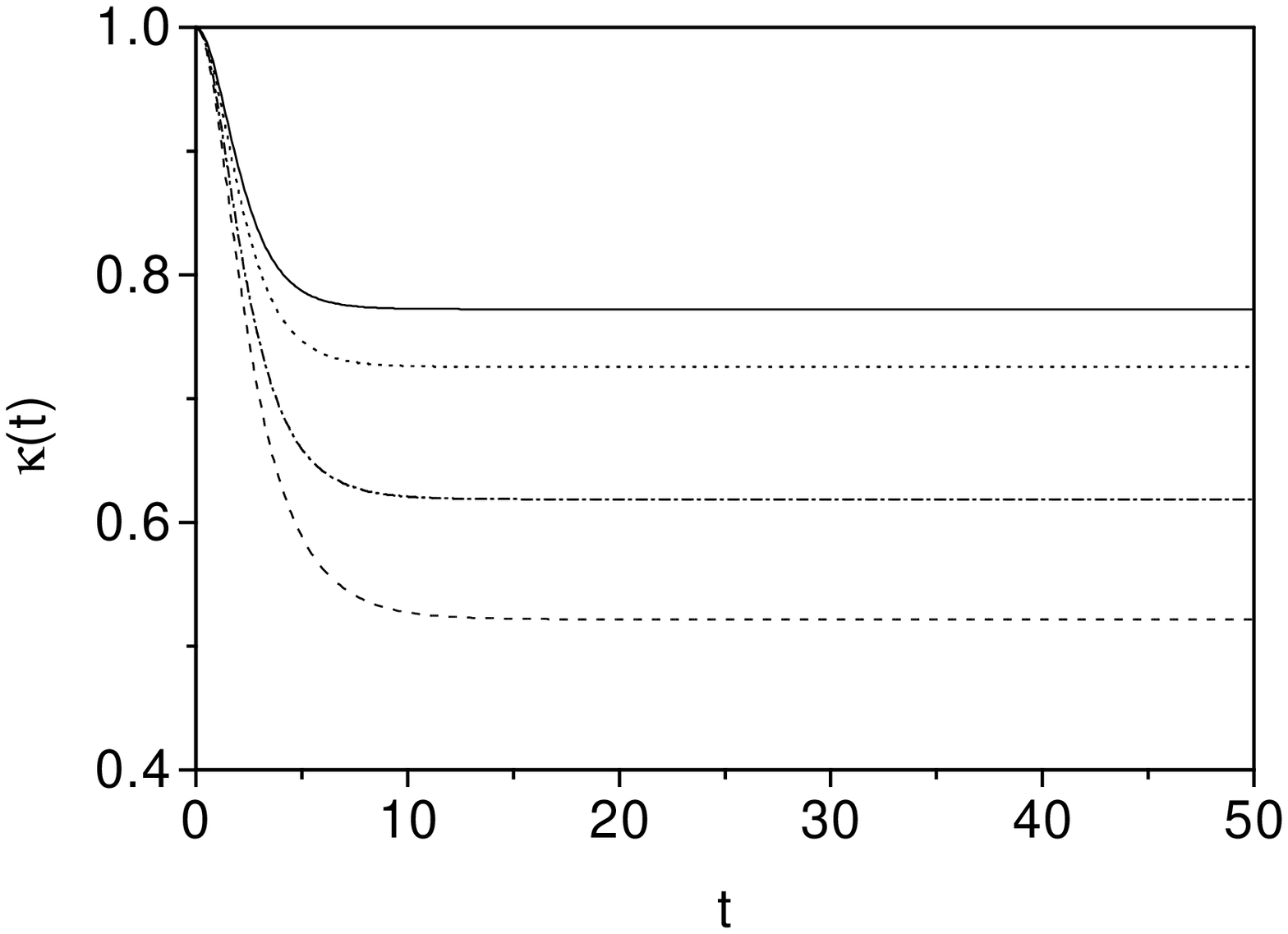}
\vspace*{0.5cm}
\caption{
The quantum transmission coefficient is plotted against time for
different values of correlation time, $\tau_c=65$ (continuous line),
$\tau_c=75$ (dotted line), $\tau_c=95$ (dashed and dotted line),
$\tau_c=110$ (dashed line) for other parameters as in Fig.4.
}
\end{figure}


\begin{figure}[h]
\includegraphics[width=0.7\linewidth,angle=-90]{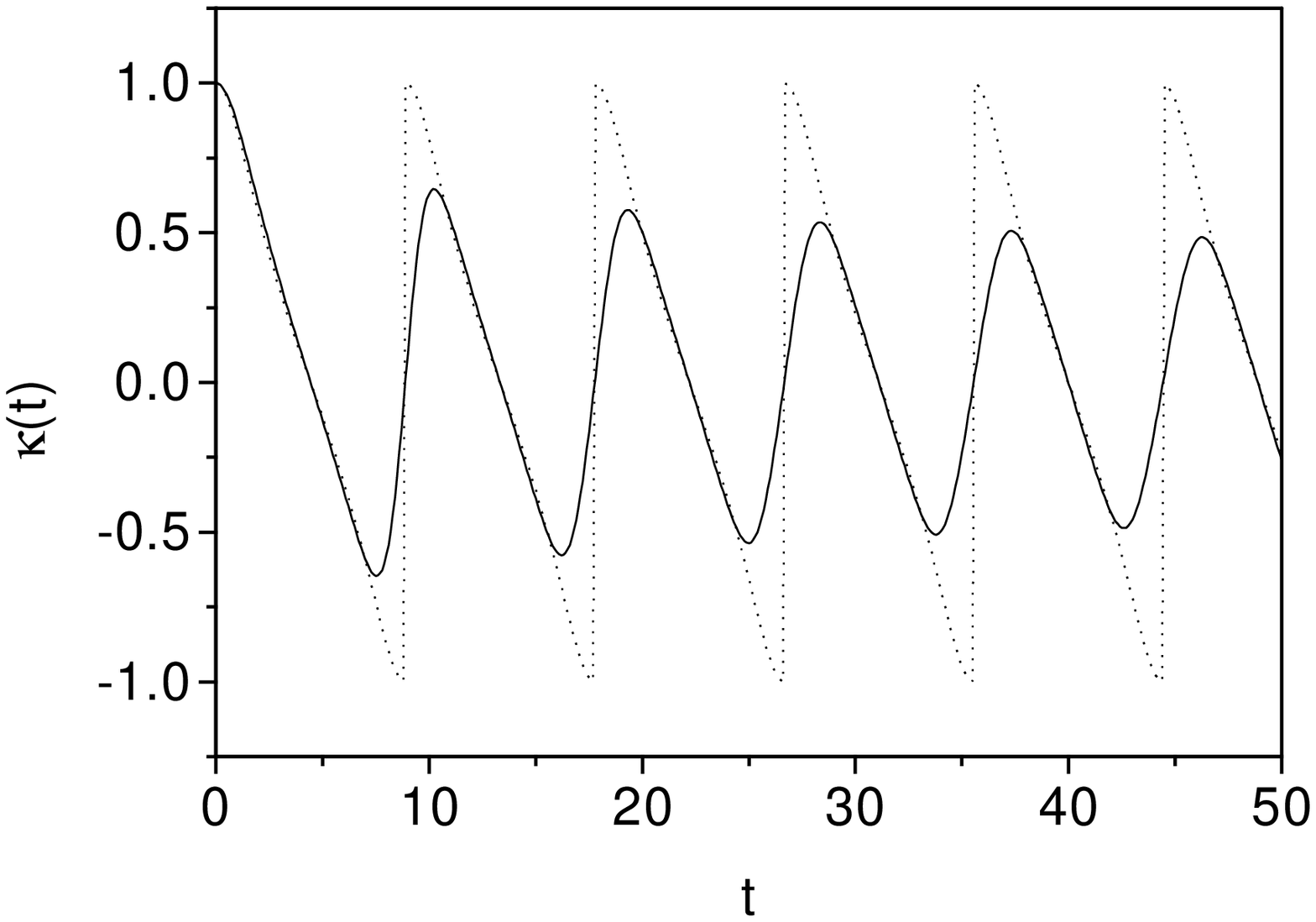}
\vspace*{0.5cm}
\caption{
The quantum (continuous line) transmission coefficient is plotted
against time and compared with classical (dotted line) one for the set of
parameters $A=150$, $\tau_c=100$, $kT=1.0$, $\omega_b=1.0$,
$\omega_0=\sqrt{2}$, $a=0.01$ (in the caging regime).
}
\end{figure}

\newpage

\begin{figure}[h]
\includegraphics[width=0.7\linewidth,angle=-90]{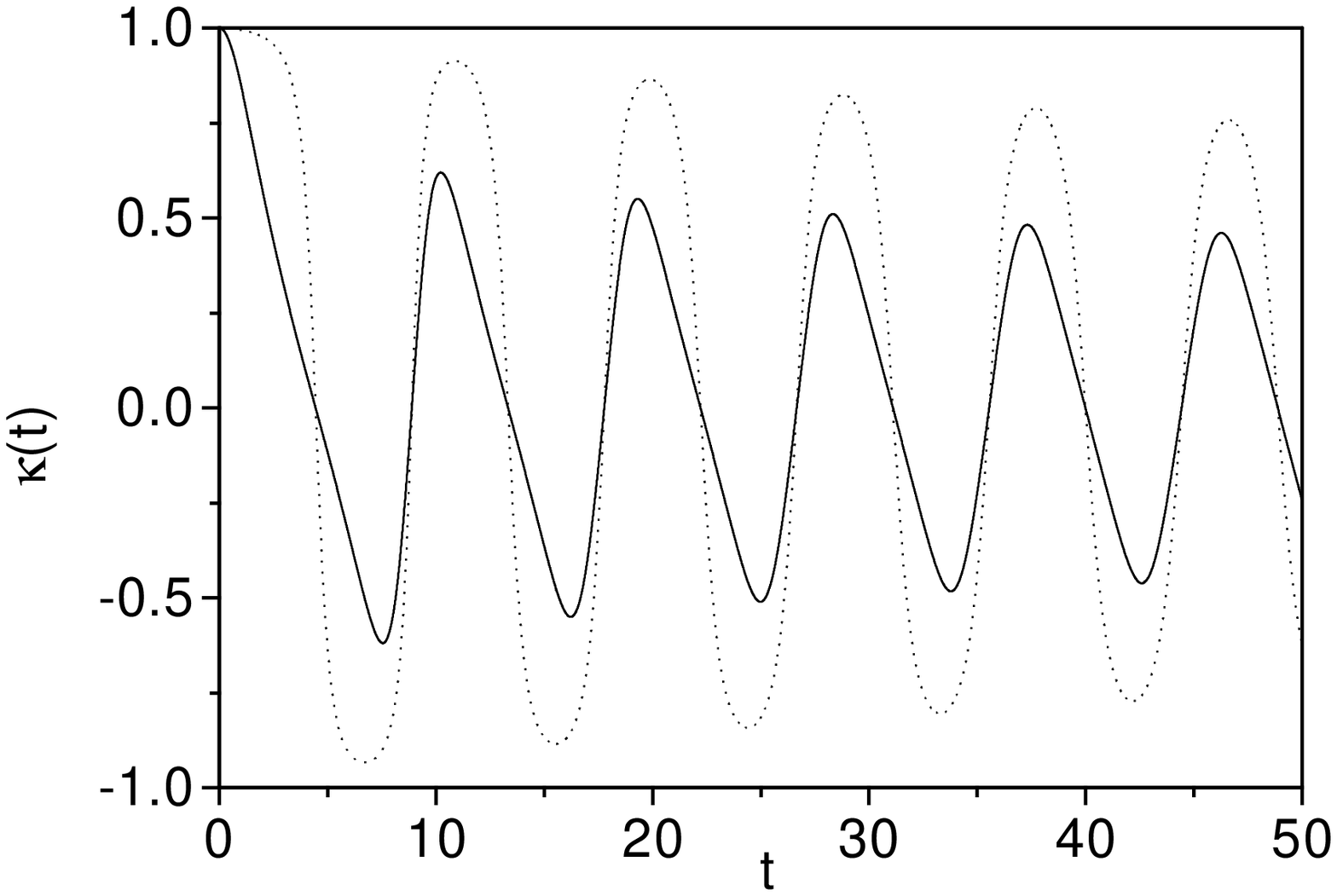}
\vspace*{0.5cm}
\caption{
Same as in Fig.7 but for $kT=0.0$ (dotted line), and $kT=25.0$
(continuous line).
}
\end{figure}


\begin{figure}[h]
\includegraphics[width=0.7\linewidth,angle=-90]{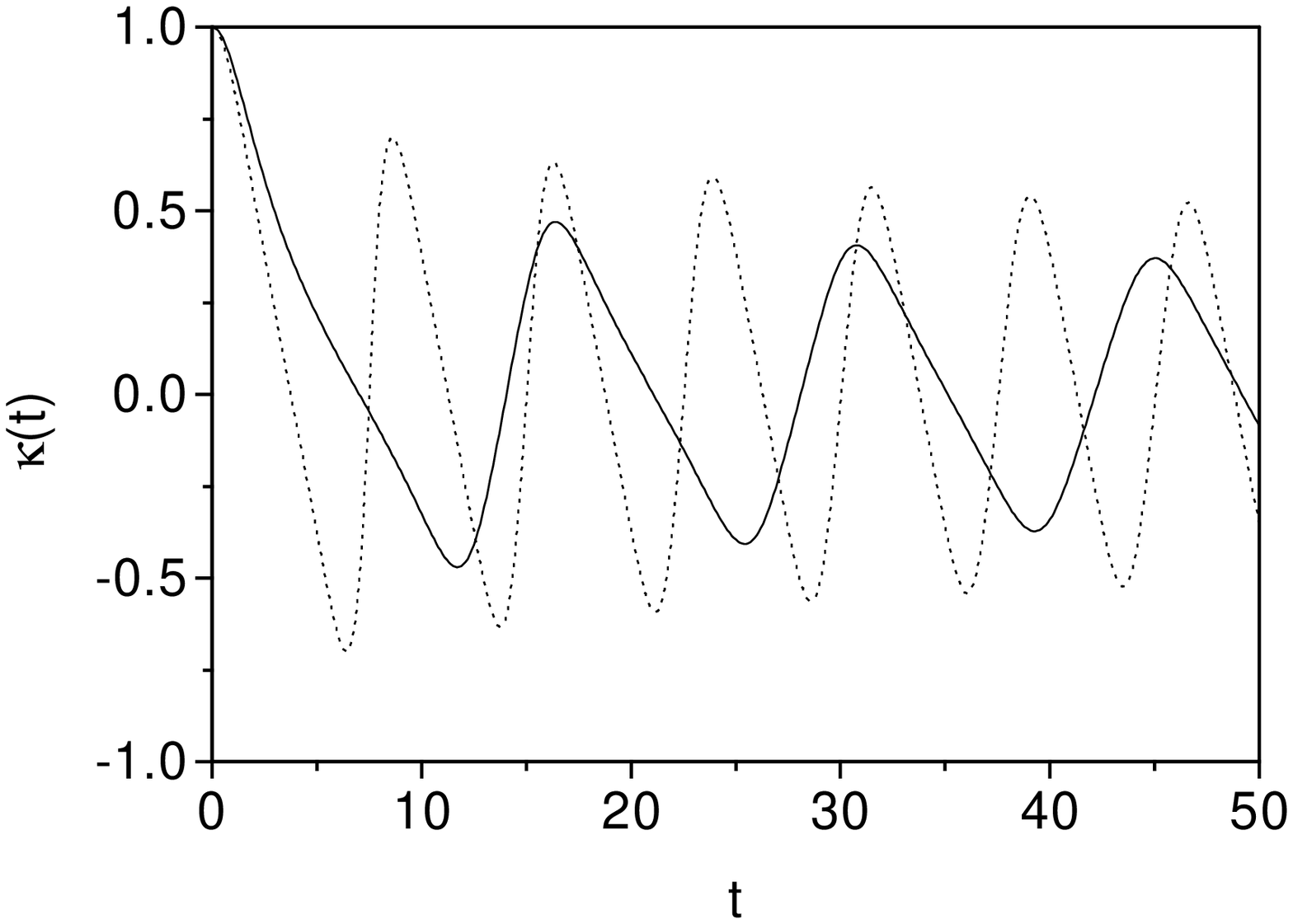}
\vspace*{0.5cm}
\caption{
Same as in Fig.7 but for $A=120$ (continuous line) and $A=170$
(dotted line).
}
\end{figure}


\begin{thebibliography}{99}

\bibitem{kramer} H.A. Kramers, Physica {\bf7}, 284 (1940).

\bibitem{dian} E.W.-G. Diau, J.L. Herek, Z.H. Kim and A.H. Zewail,
Science {\bf279}, 874 (1998).

\bibitem{caste} A.W. Castelman, D.E. Folmer and E.S.
Wisniewski, Chem. Phys. Letts. {\bf287}, 1 (1998).

\bibitem{grote-hynes} R.F. Grote and J.T. Hynes, J. Chem. Phys.
{\bf73}, 2715 (1980).

\bibitem{hang-mojta}
P. H\"anggi and F. Mojtabai, Phys. Rev. A {\bf 26}, 1168 (1982).

\bibitem{car} B. Carmeli and A. Nitzan,
J. Chem. Phys. {\bf 79}, 393 (1983).

\bibitem{lan} J.S. Langer, Ann. Phys. (N.Y.) {\bf54}, 258 (1969).

\bibitem{pollak} A.M. Berezhkovskii, E. Pollak and V. Yu
Zitserman, J. Chem. Phys. {\bf 97}, 2422 (1992).

\bibitem{gram1} R. Graham and T. Tel, Phys. Rev. Letts. {\bf 52}, 9
(1984).

\bibitem{gram2} R. Graham and T. Tel, Phys. Rev. A {\bf 31}, 1109
(1985).

\bibitem{ray1}
S.K. Banik, J. Ray Chaudhuri and D.S. Ray, J. Chem. Phys. {\bf
112}, 8330 (2000).

\bibitem{ray2}
J. Ray Chaudhuri, S.K. Banik, B.C. Bag and D.S. Ray, Phys. Rev. E
{\bf 63}, 061111 (2001).

\bibitem{ray3}
J. Ray Chaudhuri, G. Gangopadhyay and D.S. Ray, J. Chem. Phys.
{\bf 109}, 5565 (1998).

\bibitem{ray4}
J. Ray Chaudhuri, B.C. Bag and D.S. Ray, J. Chem. Phys. {\bf
111}, 10852 (1999).

\bibitem{weiss} U. Weiss, {\it Quantum Dissipative Systems},
(World Scientific, Singapore, 1999).

\bibitem{grab} H. Grabert, P. Schramm, G.L. Ingold,
Phys. Rep. {\bf 168}, 115 (1988).

\bibitem{hangg}
P. H\"anggi, P. Talkner and M. Borkovec, Rev. Mod. Phys. {\bf
62}, 251 (1990).

\bibitem{woly} P.G. Wolynes, Phys. Rev. Lett. {\bf 47}, 968 (1981).

\bibitem{mill} W.H. Miller, J. Chem. Phys. {\bf 62}, 1899 (1975).

\bibitem{calde} A.O. Caldeira and A.J. Leggett, Phys. Rev. Lett. {\bf 46}, 211
(1981).

\bibitem{grab1} H. Grabert and U. Weiss, Z. Phys. B {\bf56}, 171
(1984).

\bibitem{topa} M. Topaler and N. Makri,
J. Chem. Phys. {\bf 101}, 7500 (1994) and References given
therein.

\bibitem{bern} B.J. Berne and D. Thirumalai, Ann. Rev. Phys.
Chem. {\bf37}, 401 (1987); {\it Quantum Simulations of Condensed
Matter Phenomena edited by J. D. Doll and Gubernatis} (World
Scientific, Singapore, 1999).

\bibitem{db1} D. Banerjee, B.C. Bag, S.K. Banik and D.S. Ray,
Phys. Rev. E {\bf 65}, 021109 (2002).

\bibitem{skb} S.K. Banik, B.C. Bag and D.S. Ray,
Phys. Rev. E {\bf 65}, 051106 (2002).

\bibitem{db2} D. Banerjee, S.K. Banik, B.C. Bag, and D.S. Ray,
Phys. Rev. E {\bf 66}, 051105 (2002).

\bibitem{db3} D. Banerjee, B.C. Bag, S.K. Banik and D.S. Ray,
Physica A {\bf318}, 6 (2003)

\bibitem{bic1} D.J. Bicout, A.M. Berezhkovskii, A. Szabo and G.
H. Weiss, Phys. Rev. Letts. {\bf83} 1279 (1999); D.J. Bicout,
A.M. Berezhkovskii, and A. Szabo, J. Chem. Phys. {\bf114}, 2293
(2001).

\bibitem{sch} A. Schmid, J. Low Temp. Phys. {\bf49}, 609 (1982).

\bibitem{mag} M.O. Magnasco, Phys. Rev. Letts. {\bf71} 1477 (1993).

\bibitem{juli} R.D. Astumian, Science {\bf276}, 917 (1997);
F. Julicher, A. Adjari and J. Prost, Rev. Mod. Phys. {\bf69}, 1269
(1997); P. Reimann, Phys. Rep. {\bf 361}, 57 (2002).

\bibitem{keck} J.C. Keck, Adv. Chem. Phys. {\bf13}, 85 (1967).

\bibitem{kap} R. Kapral, J. Chem. Phys. {\bf56}, 1842 (1972).

\bibitem{chand} D. Chandler, J. Chem. Phys. {\bf68}, 2969 (1978).

\bibitem{yam} K. Yamashita and W.H. Miller, J. Chem. Phys.
{\bf82}, 5475 (1985); J.W. Tromp and W.H. Miller, Faraday Discuss
Chem. Soc. {\bf84}, 441 (1987); W.H. Miller, S.D. Schwartz and
J.W. Tromp, J. Chem. Phys. {\bf79}, 4889 (1983).

\bibitem{tan} D.J. Tannor and D. Kohen, J. Chem. Phys.
{\bf100}, 4932 (1994); D. Kohen and D.J. Tannor, {\bf103}, 6013
(1995); D. Kohen and D. J. Tannor, Adv. Chem. Phys. {\bf111}, 219
(1999).

\bibitem{san} J.M. Sancho, A.H. Romero and K. Lindenberg, J.
Chem. Phys. {\bf109}, 9888 (1998); K. Lindenberg, A.H. Romero and
J.M. Sancho, Physica D {\bf133}, 348 (1999).

\bibitem{loui}
W. H. Louisell, {\it Quantum Statistical Properties of Radiation}
(Wiley, New York, 1973).

\bibitem{sm} B. Sundaram and P.W. Milonni,
Phys. Rev. E {\bf 51} 1971, (1995)  .

\bibitem{akp} A.K. Pattanayak and W.C. Schieve,
Phys. Rev. E {\bf 50} 3601, (1994).

\bibitem{talk} P. Talkner, Chem. Phys. {\bf180}, 199 (1994).

\bibitem{melni} V. I. Melnikov, Phys. Rev. E {\bf 48}, 3271 (1993).

\bibitem{voth} E. Geva, Q. Shi and V. A. Voth, J. Chem. Phys.
{\bf115} 9209, (2002).

\bibitem{wsmi} H. Wang, X. Sun and W. H. Miller, J. Chem. Phys.
{\bf108}, 9726 (1998).

\bibitem{leg} A. O. Caldeira and A. J. Leggett, Ann. Phys. {\bf149}, 374 (1983).

\bibitem{bol} A. O. Bolivar, Phys. Rev. A, {\bf58}, 4330 (1998).

\bibitem{boli} A. O. Bolivar, Physcia A {\bf301}, 219 (2001).

\bibitem{lsf} L. S. F. Olavo, Physcia A {\bf262}, 197 (1999).

\bibitem{jes} J. E. Staub and B. J. Berne, J. Chem. Phys. {\bf83},
1138 (1985).

\bibitem{jest} J. E. Staub, D. A. Hsu and B. J. Berne, J. Phys.
Chem. {\bf89}, 5188 (1985).

\bibitem{dsr} D. S. Ray,J. Chem. Phys. {\bf92}, 1145 (1990).

\bibitem{rad} J. M. Radcliffe, J. Phys. A. {\bf4}, 313 (1971)

\end{thebibliography}
\end{document}